\documentclass[12pt,draftclsnofoot,onecolumn]{IEEEtran}
\usepackage{blindtext}
\usepackage{graphicx}
\usepackage{graphicx,subfigure}
\usepackage{array}
\usepackage{url}
\usepackage{algorithmic}
\usepackage[cmex10]{amsmath}
\usepackage{ifpdf}
\usepackage{amssymb,amsmath}
\usepackage{empheq}
\usepackage{stfloats}   
\usepackage{cite}
\usepackage{algorithm}
\usepackage{algorithmic}
\usepackage{multirow}
\usepackage{longtable}

\usepackage[dvipsnames,usenames]{color}
\usepackage{prettyref}
\usepackage[normalem]{ulem}
\usepackage{amsthm}

\theoremstyle{remark}

\theoremstyle{remark}

\theoremstyle{remark}
\newtheorem{rmk}{Remark}
\usepackage{graphicx}
\usepackage{ifpdf}
\usepackage{cite}
\usepackage{xcolor}
\usepackage{enumerate}
\usepackage{bbm}
\newcommand{\figref}[1]{\figurename~\ref{#1}}

\usepackage{array}
\usepackage{mdwmath}
\usepackage{mdwtab}
\usepackage{eqparbox}
\usepackage{bm}

\begin{document}

\title{CNN-Based Signal Detection for Banded Linear Systems}
 
\author{Congmin~Fan,~Xiaojun~Yuan,~\IEEEmembership{Senior Member,~IEEE,} and~Ying-Jun~Angela~Zhang,~\IEEEmembership{Senior~Member,~IEEE}
\thanks{The work in this paper will be partially presented in IEEE Globecom 2018 \cite{fan2018deep}.}
}
\maketitle
\begin{abstract} 
Banded linear systems arise in many communication scenarios, e.g., those involving inter-carrier interference and inter-symbol interference. Motivated by recent advances in deep learning, we propose to design a high-accuracy low-complexity signal detector for banded linear systems based on convolutional neural networks (CNNs). We develop a novel CNN-based detector by utilizing the banded structure of the channel matrix. Specifically, the proposed CNN-based detector consists of three modules: the input preprocessing module, the CNN module, and the output postprocessing module. With such an architecture, the proposed CNN-based detector is adaptive to different system sizes, and can overcome the curse of dimensionality, which is a ubiquitous challenge in deep learning. Through extensive numerical experiments, we demonstrate that the proposed CNN-based detector outperforms conventional deep neural networks and existing model-based detectors in both accuracy and computational time. Moreover, we show that CNN is flexible for systems with large sizes or wide bands. We also show that the proposed CNN-based detector can be easily extended to near-banded systems such as doubly selective orthogonal frequency division multiplexing (OFDM) systems and 2-D magnetic recording (TDMR) systems, in which the channel matrices do not have a strictly banded structure. 
\end{abstract}
\newpage
\section{Introduction}
\subsection{Background and Motivations}
Detection of modulated signals based on noisy channel observations in the presence of interference is one of the most basic building blocks in communication systems. It has been a long-standing challenge to design a high-accuracy and low-complexity signal detection method that performs well for general communication systems. Intensive research endeavors have been focused on exploiting special structures of communication systems to design efficient signal detection methods. For example, the channel sparsity in massive multiple-input multiple-output (MIMO) systems \cite{zhang2017blind} and cloud radio access networks \cite{fan2017scalable} has been utilized to design message-passing-based detection algorithms with low complexity. Liu \textit{et al.} proposed a discrete first-order detection method for large-scale MIMO detection with provable guarantees based on the independent and identically (i.i.d.) distributed channel coefficients \cite{liu2017discrete}. In this paper, we focus on banded linear systems, in which the channel matrices are banded matrices. The banded structure of a system can be caused by, e.g., inter-carrier interference (ICI) and inter-symbol interference (ISI). For instance, in frequency selective channels, ISI arises between adjacent received symbols, yielding a banded channel matrix \cite{leus2011estimation}. Similarly, 2-D magnetic recording (TDMR) systems typically suffer from 2-D banded ISI caused by a combination of down-track ISI and intertrack interference at the read head \cite{carosino2015iterative}. In doubly selective channels, orthogonal frequency division multiplexing (OFDM) systems may experience significant ICI from adjacent subcarriers, which implies that the channel matrix in the frequency domain can be approximated as a banded matrix \cite{liu2015banded}. Traditional detectors ignoring the banded structure will lead to inferior performance. For example, detectors designed for interference-free systems will cause a low estimation accuracy. Meanwhile, detectors designed for general interference systems usually have very high computational complexity.

The banded structure of the channel has been extensively studied to reduce the complexity in signal detection. For example, the well-known Bahl-Cocke-Jelinek-Raviv (BCJR) algorithm \cite{cocke1974optimal} can be employed in a banded system to achieve the optimal maximum \textit{a posteriori} probability (MAP) detection. Nevertheless, this approach is disadvantageous in communication systems with large signal dimensions due to its intrinsic serial algorithm and exponential complexity in band width. In \cite{rugini2005simple}, Rugini \textit{et al.} proposed to reduce the complexity of the linear maximum mean square error (LMMSE) detector through LDL$^H$ factorization. However, there exists a considerable performance gap between the linear detector and the MAP detector. Iterative algorithms including iterative MMSE \cite{schniter2004low} and belief propagation \cite{ochandiano2011iterative} have been proposed as near-optimal solutions. These iterative algorithms typically require a large number of iterations to obtain an estimate with high accuracy. Moreover, it is difficult to efficiently implement the iterative algorithms in parallel, which significantly limits the computational efficiency. In a nutshell, there is a fundamental tradeoff between computational complexity and detection accuracy in signal detection problems. It is highly desirable to design a detection algorithm that achieves both high accuracy and low complexity for banded linear systems, which is the focus of this paper.

\subsection{Contributions}
Motivated by the recent advances in deep learning \cite{lecun2015deep}, we aim to design high-accuracy low-complexity signal detectors based on deep neural networks (DNNs). 
Instead of using a general DNN, we propose to design the detector based on a convolutional neural network (CNN) that consists of only convolutional layers. The reasons for CNN-based signal detection are explained as follows. First, it is well known that DNNs with fully connected layers suffer from the curse of dimensionality, i.e., the number of tunable parameters significantly grows as the system size increases. In a CNN, all neurons in a layer share the same set of tunable parameters, which addresses the curse of dimensionality. Secondly, a DNN with fully-connected layers has to be retrained once the system size changes. In contrast, when the tunable parameters are well-trained, a CNN can be applied to systems with different sizes without the need of retraining.

Despite the advantages of being scalable and robust to the system size, it is nontrivial to employ CNN for signal detection. The success of CNN is based on the assumption that if one set of parameters is useful to extract a feature at a certain spatial position, then the same set of parameters is also useful to extract the feature at other positions. Such shift-invariance assumption, although holds in many computer vision problems, does not hold in a signal detection problem. To address this challenge, we propose a novel CNN-based detection architecture consisting of three modules: an input preprocessing module, a CNN module, and an output postprocessing module. The input preprocessing module reorganizes the input (i.e., the channel matrix and the received signals in this paper) based on the banded structure to obtain the shift-invariance property. Then, the shift-invariant input is fed into the CNN, the output of which is processed through the output postprocessing module to give an estimate of the transmitted signals. To the best of our knowledge, our work is the first attempt to design a CNN-based detector for banded linear systems. 

We conduct extensive numerical experiments to show that the CNN-based detector performs much better than existing detectors with comparable complexity. Moreover, the proposed CNN demonstrates outstanding robustness for different system sizes. It achieves a high accuracy even if there is a mismatch between the system sizes in the training set and the testing set. In addition, we extend the proposed CNN-based detector to near-banded channels, such as 1-D near-banded channels in doubly selective OFDM systems and 2-D near-banded channels in TDMR systems with 2-D ISI. Specifically, we propose a cyclic CNN (CCNN) for 1-D near-banded channels, and propose a 2-D CNN-based detector for 2-D near-banded channels. Through simulations, we show that the proposed detector still performs well in these systems, where the channel matrix is not in a strictly banded structure. 

In summary, the benefit of the proposed CNN-based detector is at least fourfold. 
\begin{itemize}
    \item The proposed CNN approach relieves the burden to establish a sophistical mathematical model for the communication system, since it provides a universal detector that automatically adapts to any channel and noise distributions.
    \item The CNN-based detector achieves much better error performance than the other detectors with comparable computational complexity, and is ideally constructed for parallel computing.
    \item Thanks to the parameter-sharing property, the proposed CNN is robust to mismatched system sizes in the training set and the testing set.
    \item The CNN-based detector can be readily extended to systems without a strictly banded structure. As such, the proposed CNN approach sheds lights on how to design CNN-based algorithms for other problems in communication systems with a near banded structure. 
\end{itemize}

\subsection{Related Work}
Recently, there have been two threads of research on the application of deep learning for signal detection in communication systems. The first thread is to design deep learning based detectors by unfolding existing iterative detection algorithms. That is, the iterations of the original algorithm are unfolded into a DNN with each iteration being mimicked by a layer of the neural network. Instead of predetermined by the communication model (i.e., the channel matrix, the modulation scheme, the distribution of noise, etc.), the updating rule at each layer is controlled by some tunable parameters, which are learned based on the training data. For example, \cite{gregor2010learning} unfolded two well-known algorithms, namely iterative shrinkage and thresholding algorithm (ISTA) \cite{beck2009fast} and approximate message passing (AMP) \cite{rangan2011generalized}, for a fixed channel matrix. It is shown that the proposed neural networks significantly outperform the original algorithms in both computational time and accuracy \cite{gregor2010learning}. The second thread is to treat the transmission procedure as a black box, and utilize conventional DNNs for signal detection. \cite{ye2018power} showed that a fully connected neural network is able to detect signals for various channel realizations. Specifically, \cite{ye2018power} utilized deep learning to realize joint channel estimation and signal detection in OFDM systems, where the channel matrix is diagonal. It is demonstrated that the deep learning approach achieves a higher detection accuracy than existing model-based detection approaches with comparable complexity. In \cite{farsad2018neural}, Farsad \textit{et al.} presented a recurrent neural network (RNN) for detection of data sequences in a Poisson channel model, which is applicable to both optical and chemical communication systems. The proposed RNN can achieve a performance close to the Viterbi detector with perfect CSI. 

Besides signal detection, deep learning has demonstrated its potential in other areas of communication systems. Nachmani \textit{et al.} studied the problem of channel decoding through unfolding traditional belief propagation (BP) decoders \cite{nachmani2016learning}. Most recently, Liang \textit{et al.} proposed an iterative belief propagation-CNN architecture for channel decoding under a certain noise correlated model \cite{liang2018iterative}. A standard BP decoder is used to estimate the coded bits, followed by a CNN to remove the estimation errors of the BP decoder, and obtain a more accurate estimation. In \cite{dorner2017deep}, Dorner \textit{et al.} presented an end-to-end communication system to demonstrate the feasibility of over-the-air communication with deep neural networks. As shown in \cite{dorner2017deep}, the performance is comparable with traditional model-based communication systems.

\subsection{Organization}
The rest of the paper is organized as follows. In Section II, we present the system model as well as its extensions to near-banded systems, and discuss the challenges of utilizing traditional DNNs to detect signals. In Section III, we propose the CNN-based detector based on the banded structure of the channel matrix, and illustrate the robustness of the proposed detector. In Section IV, we extend the proposed detector to near-banded systems. In Section V, the performance of the proposed deep learning approach is evaluated in different channel models, and is compared with existing algorithms. In Section V, we also show the performance of the proposed CNN in practical OFDM systems and TDMR systems. Conclusions and future work are presented in Section VI.
\begin{figure}[!h]
\centering
\includegraphics[scale=0.45]{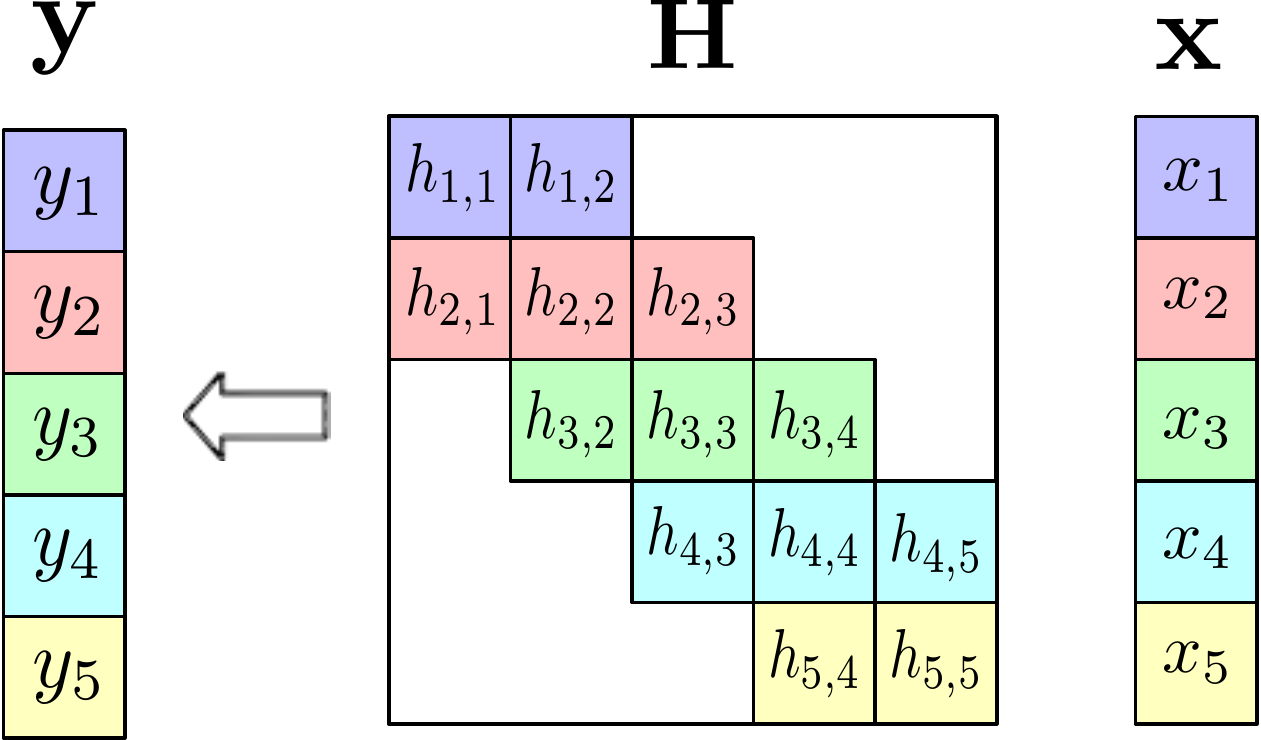}
\caption{A linear banded system with $K = 5$ and $B=1$.}\label{fig:channel1}
\end{figure}
\section{System Model}
\subsection{Linear Banded Systems}
In this paper, we consider a linear channel model with the received signal $\mathbf {y}\in \mathbb{C}^{K}$ written as
\begin{equation}
    \mathbf{y}=\mathbf{H}\mathbf{x}+\mathbf{n}, \label{eqn:y}
\end{equation}
where $\mathbf{H}\in \mathbb{C}^{K\times K}$ is the channel matrix, $\mathbf{x}\in \{\pm 1\}^{K}$ is the vector of transmitted signals\footnote{For simplicity, we use BPSK as the modulation method, but the proposed deep learning approach can be readily extended to systems with other modulation methods.}, and $\mathbf{n} \in \mathbb{C}^{K}$ is the noise vector. Furthermore, we assume that the channel matrix is a banded matrix with bandwidth $B$. That is, 
\begin{equation}
    H_{k,m} = 0, \text{if } |k-m|>B,
\end{equation}
where $H_{k,m}$ is the $(k,m)$th element in the channel matrix $\mathbf{H}$ and $B$ is the bandwidth of the channel matrix (see \figref{fig:channel1}). Under this assumption, the $k$th entry of $\mathbf y$ in (\ref{eqn:y}) can be rewritten as
\begin{equation}
    y_k=\sum_{b=-B}^B H_{k,k+b}x_{k+b} + n_k,\label{eqn:band}
\end{equation}
where $x_k$ and $n_k$ are the $k$th entries of $\mathbf{x}$ and $\mathbf{n}$, respectively.\footnote{In (\ref{eqn:band}), we assume $H_{k,k+b}=0$ for $k+b\leq 0$ and $k+b>K$.} We assume perfect channel state information at the receiver, i.e., the channel matrix $\mathbf{H}$ is exactly known by the receiver.

The banded system in (\ref{eqn:band}) may be idealized in practical scenarios. We next introduce two near-banded systems with the channel matrices obtained from real applications. We will show that the CNN-based detector can be readily modified to handle the near-banded systems.

\begin{figure}[!h]
\centering
\includegraphics[scale=0.45]{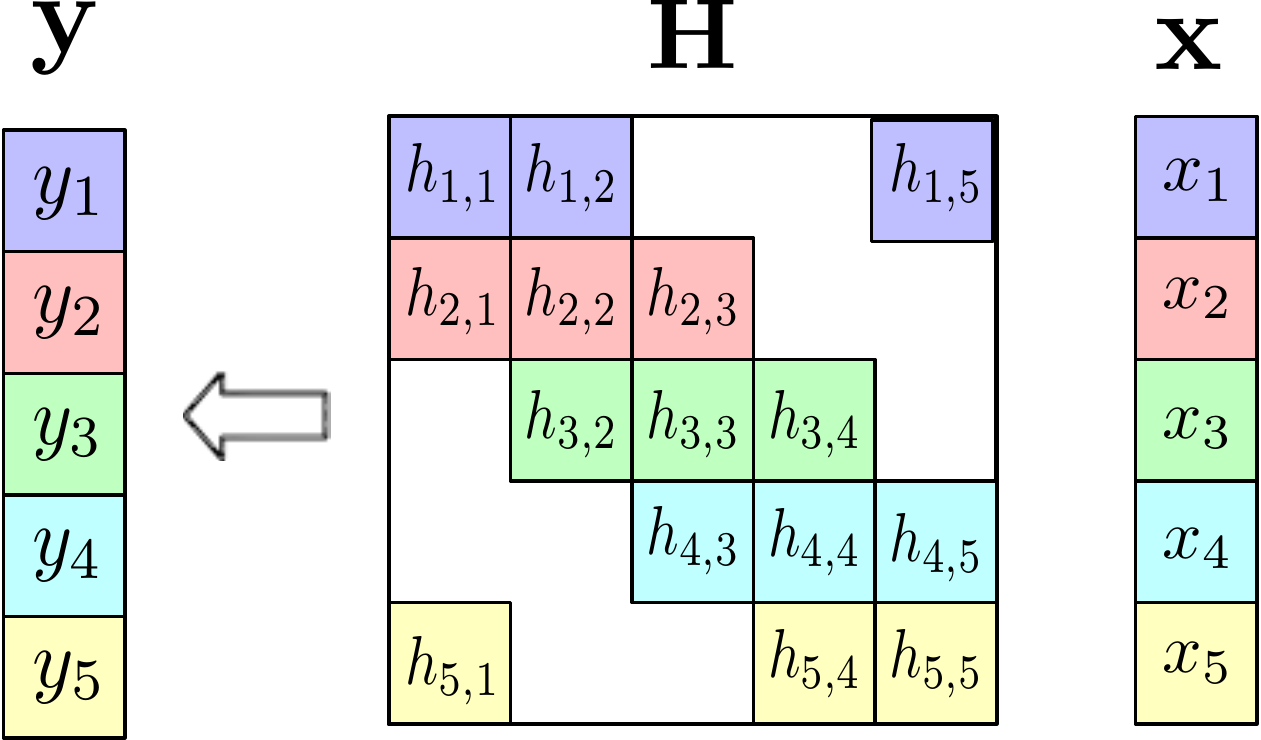}
\caption{A doubly selective OFDM system with $K = 5$ and $B=1$.}\label{fig:channel2}
\end{figure}
\begin{figure}[!h]
\centering
\includegraphics[scale=0.45]{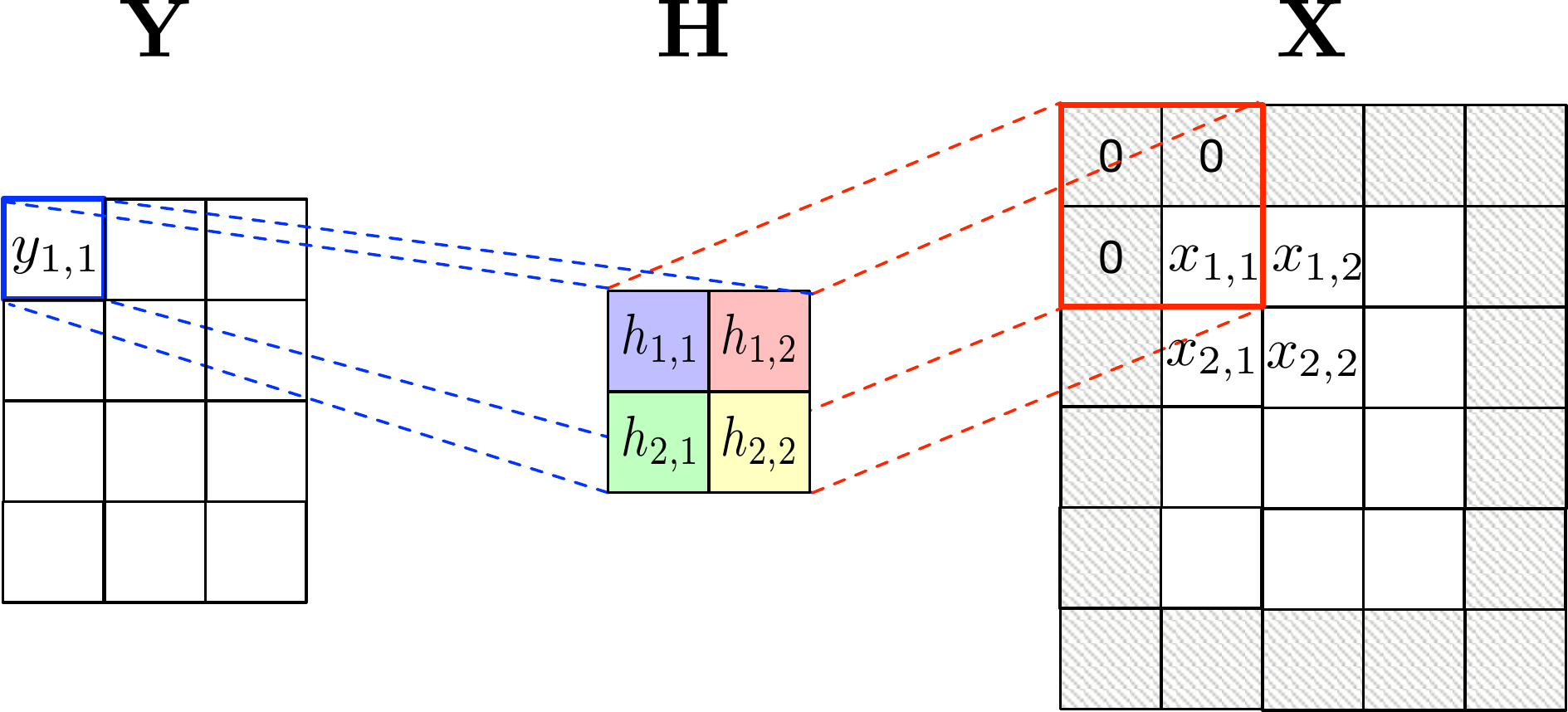}
\caption{A TDMR system with $N = 4$, $K =3$, and $B = 1$.}\label{fig:channel3}
\end{figure}
\subsection{Near-Banded Systems}
\subsubsection{1-D Near-Banded systems}
In certain systems, such as systems with a doubly selective OFDM channel, in addition to the non-zero entries on the diagonal band, the channel matrix has non-zero entries in the bottom-left corner and the top-right corner due to the non-negligible ICI \cite{schniter2004low}. The structure of the channel matrix is shown in \figref{fig:channel2}, where the entries of the channel matrix satisfy
\begin{equation}
    H_{k,m} = 0, \text{if } B<|k-m|<K-B.\label{eqn:nearband}
\end{equation}
\subsubsection{2-D Near-Banded systems}
A TDMR system usually suffers from 2-D banded ISI modeled by convolving the data with a 2-D spatial impulse response \cite{wu2003iterative}. The output of the channel is a matrix $\mathbf{Y} \in \mathbb{R}^{N\times K}$ with the $(n,k)$-th element given by
\begin{equation}
    y_{n,k} = \sum_{m=1}^{B+1}\sum_{l=1}^{B+1} x_{n-m-1,k-l-1}h_{m,l}+n_{n,k}, \label{eqn:2D-ISI}
\end{equation}
where $n_{n,k} \in \mathbb{R}$ is the noise, $\mathbf{h} \in \mathbb{R}^{(B+1)\times (B+1)}$ is a 2-D read head impulse response, and $B$ is the number of elements over which the ISI extends in each dimension. As shown in \figref{fig:channel3}, the TDMR ISI system is actually a 2-D extension of the banded linear system. That is, each received signal in a TDMR system is a linear combination of the neighbouring transmitted signals in the 2-D space. 

Signal detection in a near-banded system is usually more challenging due to the more complicated structure of the interference. As shown in Section IV, the proposed CNN-based detector can be readily extended to these near-banded systems, and hence is more flexible than traditional model-based detectors.

\begin{figure}[!h]
\centering
\includegraphics[scale=0.45]{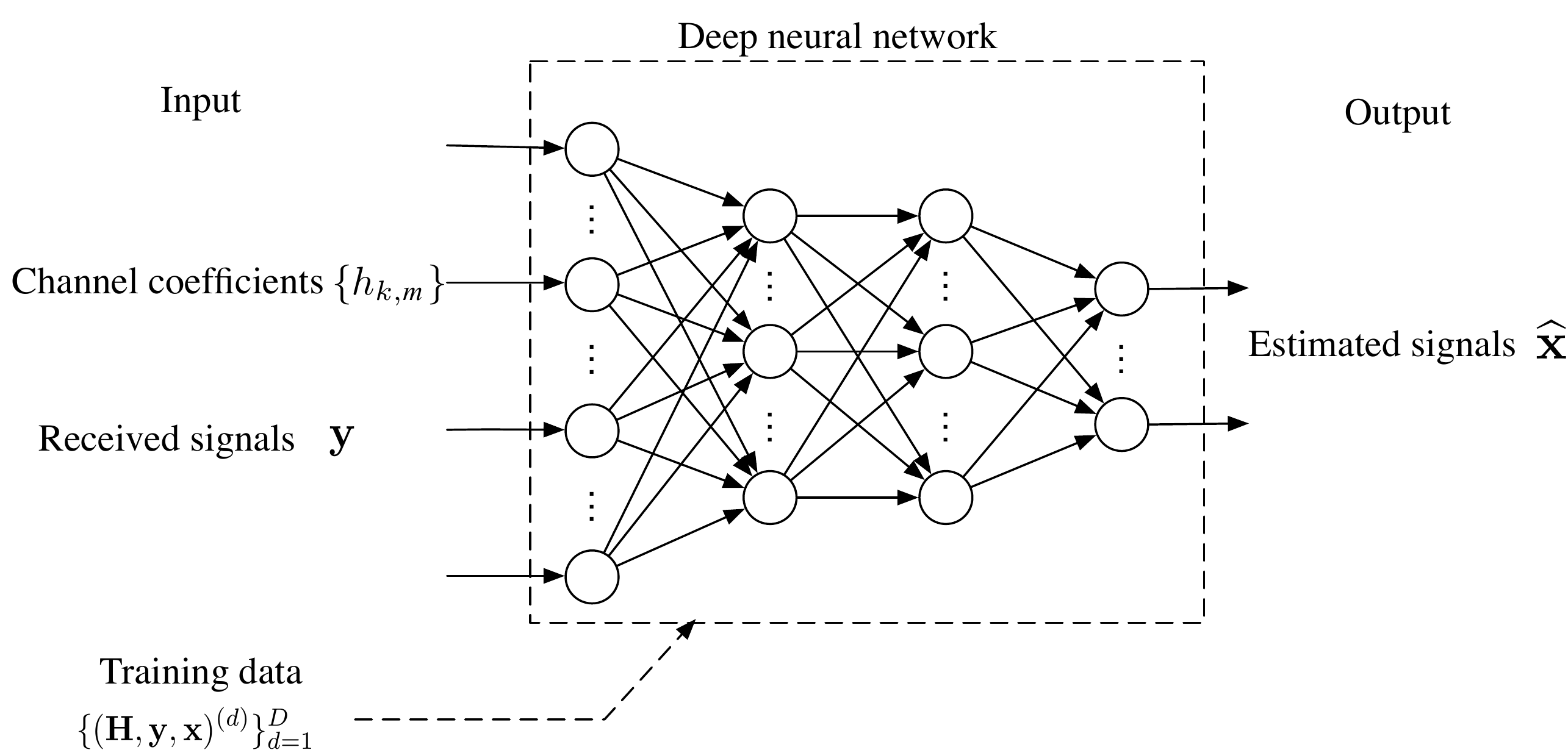}
\caption{Architecture of a DNN-based detector.}\label{fig:DNN}
\end{figure}
\subsection{Architecture of a DNN-Based Detector}
In this subsection, we briefly introduce the architecture of a DNN-based detector. As shown in \figref{fig:DNN}, the DNN-based detector treats both the channel matrix $\mathbf H$ and the received signal $\mathbf y$ as input and outputs a vector of estimated symbols $\widehat{\mathbf{x}}$. This implies that once well-trained, the proposed DNN-based detector can adapt to various channel realizations. Moreover, unlike most existing detection approaches based on the probability model of the system in (\ref{eqn:y}), the DNN based approach does not rely on the probability distributions of the channel coefficients and the noise $\mathbf{n}$. Instead, the proposed neural networks are able to learn the model information from the training data. 

Typically, a DNN may consist of fully-connected layers, densely-connected layers, convolutional layers, or their mixture. Due to the huge amount of connections between neurons, a DNN with fully-connected or densely-connected layers suffers from the curse of dimensionality, and does not scale well to large systems. More specifically, the number of weights and biases associated with each fully-connected or densely-connected neuron grows linearly with the size of the input. This means that the total number of tunable parameters increases quadratically with the size of input, which renders it difficult to train a DNN for a large system. Moreover, a DNN has to be retrained once the system size changes, because the number of tunable parameters varies with the system size. Noticeably, the DNN training is a time-consuming task, as it usually involves a large amount of data and requires high computational complexity. To deal with these challenges, we propose to detect signals through a DNN that consists of only convolutional layers (or called CNN). In a CNN, all neurons in a layer share the same set of tunable parameters, implying that the number of tunable parameters does not scale with the system size. Nonetheless, to achieve good performance with a CNN, the input is required to have shift-invariant properties, and the convolutional filter is required to be carefully designed. In the next section, we introduce the proposed CNN-based detector for strictly banded linear systems. The extension to near-banded systems will be discussed in Section IV.

\section{CNN-Based Detector}
In this section, we first describe the design details of the CNN-based detector. Then, we demonstrate the robustness of the proposed detector in the sense of adapting to various system sizes.

\begin{figure}[!h]
\centering
\includegraphics[scale=0.45]{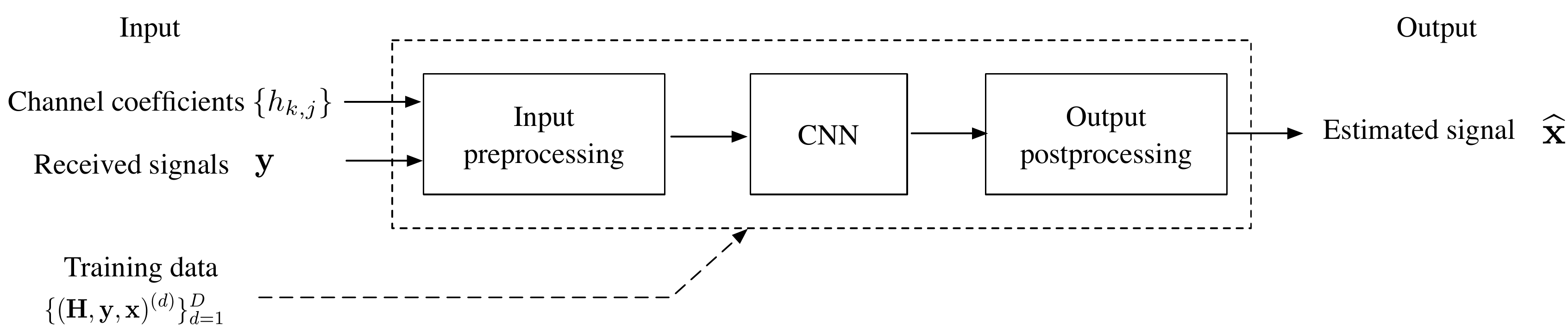}
\caption{Architecture of the CNN-based detector.}\label{fig:CNN_new}
\end{figure}
\begin{figure*}[!h]
\centering
\subfigure[Input preprocessing]{
\includegraphics[scale=0.42]{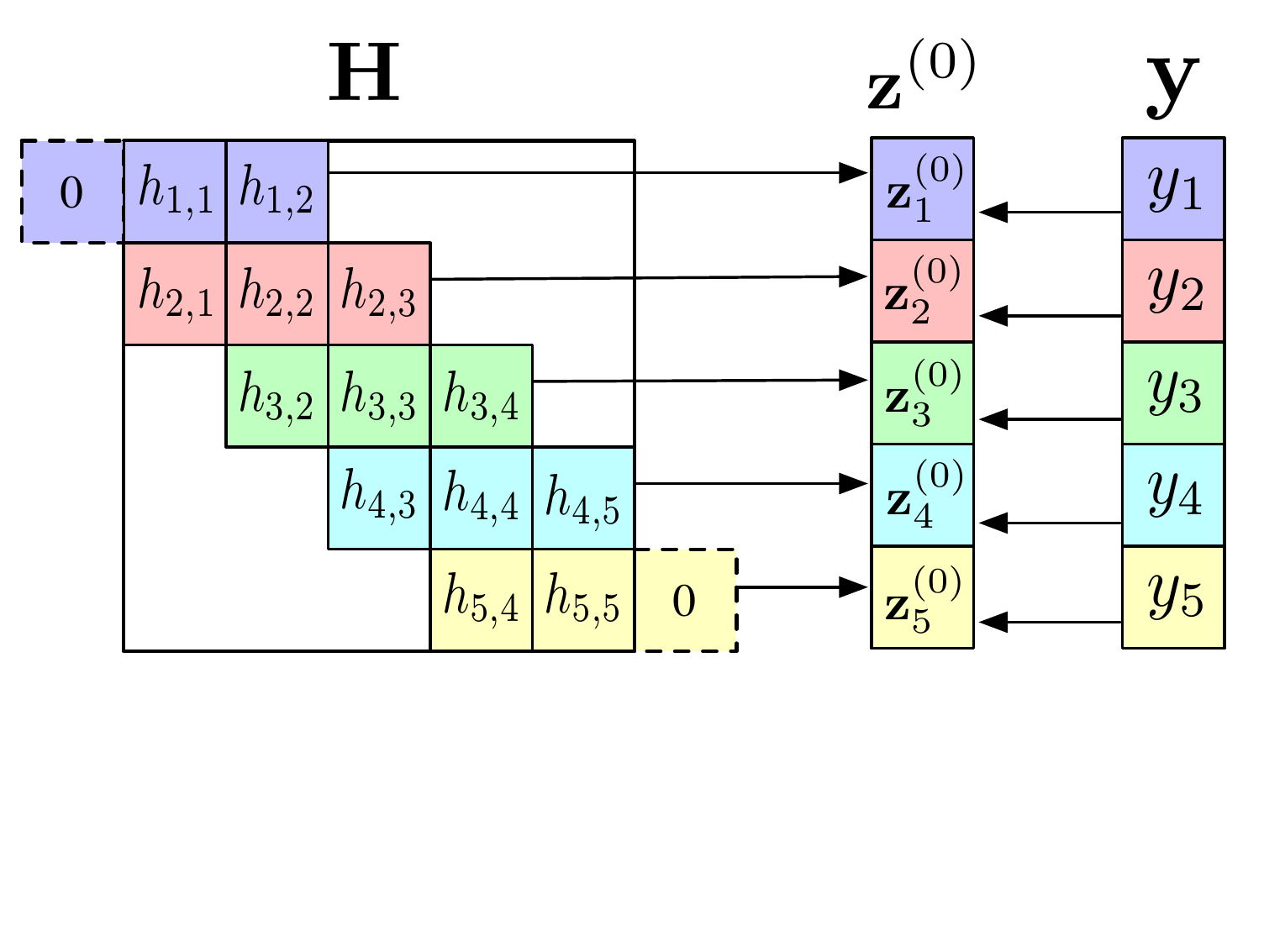}}\label{fig:CNN_1}
\subfigure[CNN architecture]{
\includegraphics[scale=0.42]{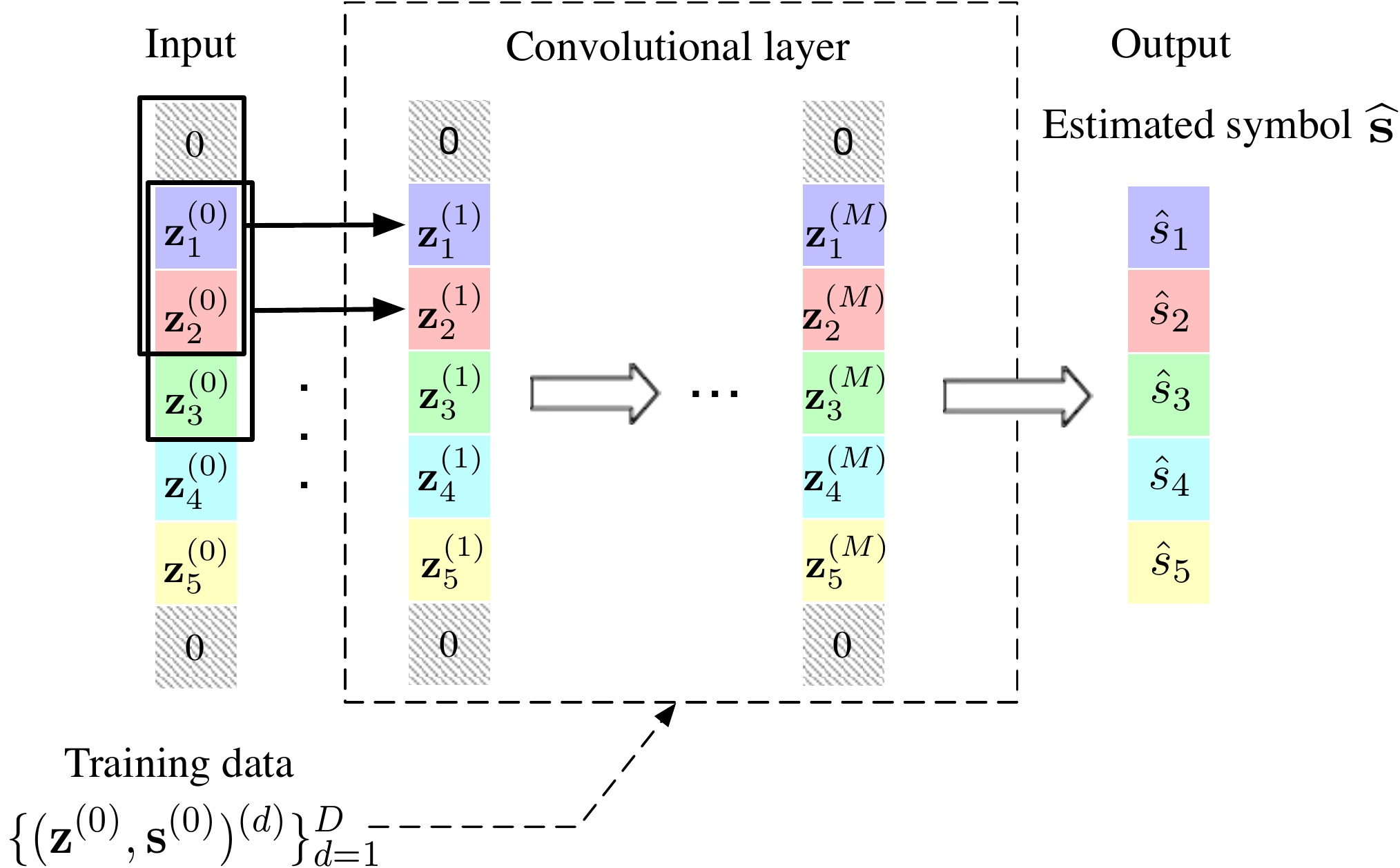}}\label{fig:CNN_2}
\caption{The convolutional neural network with $M$ convolutional layers for a system with $K=5$ and $B=1$.} \label{fig:CNN}
\end{figure*}

To address the challenges discussed in Section II, we propose to use a CNN consisting of only convolutional layers for signal detection in a banded linear system. In a convolutional layer, each neuron is only connected to a small portion of neurons in the previous layers, and all neurons in a layer share the same set of parameters (i.e., weights and biases). This significantly reduces the total number of parameters in learning. CNN is a very efficient class of DNNs for solving problems with a large-sized input, such as image/video recognition \cite{he2016deep}, natural language processing \cite{kim2014convolutional}, speech recognition \cite{abdel2014convolutional}, etc. Nonetheless, the success of CNN is based on the shift-invariance assumption. That is, if one set of parameters is useful to extract a feature at a certain spatial and temporal position, then it is also useful to extract the feature at other positions. Such assumption generally holds in image, video, and audio inputs. However, it does not necessarily hold in the signal detection problem over the channel in (\ref{eqn:y}). For example, directly shifting the channel matrix will significantly change the transmission model and thereby change the detection result. Hence, in the proposed CNN-based detector, the input as well as the tunable convolutional filter needs to be appropriately organized before fed into the CNN. As illustrated in \figref{fig:CNN_new}, we propose a CNN-based detector that consists of three modules: an input preprossing module, an CNN module, and an output postprocessing module. The input preprocessing module is used to reorganize the input to ensure the shift-invariance property. The CNN module is a CNN to extract the features from the shift-invariant input. The output postprocessing module is applied to obtain an estimate of the transmitted signals based on the features extracted by the CNN. In the following subsections, we will discuss the detailed design of the three modules.

\subsection{Input Preprocessing}
In the input preprocessing module, we use an input reshaping approach to ensure the shift-invariance property of the input $\{h_{k,m}\}$ and $\mathbf{y}$.\footnote{The realization of the input preprocessing module to achieve shift-invariance is not unique. The input reshaping approach proposed in this paper is just an example.}

As illustrated in \figref{fig:CNN}(a), we reshape the channel coefficients and the received signals into a vector $\mathbf{z}^{(0)}$. Recall that the channel matrix is a banded matrix, with the non-zero entries confined to a diagonal band. Hence, we only need to store the non-zero entries on the band into the vector $\mathbf{z}^{(0)}$. Specifically, the non-zero channel coefficients and the received signal corresponding to receiving position $k$ are stored in a vector $\mathbf{z}_k^{(0)}$ with the entries given by
\begin{equation}
    z_{k}^{(0)} [2i+1]\!=\! \begin{cases}
    \text{Re}(H_{k,k-B+i}), &i\leq 2B, 1\leq k\!-\!B\!+\!i \leq K, \\
    \text{Re}(y_{k}), & i=2B+1,\\
    0, &\text{otherwise},
    \end{cases}
\end{equation}
and
\begin{equation}
    z_{k}^{(0)} [2i+2]\!=\!\begin{cases}
    \text{Im}(H_{k,k-B+i}), &i\leq 2B, 1\leq k\!-\!B\!+\!i \leq K, \\
    \text{Im}(y_{k}), & i=2B+1,\\
    0, &\text{otherwise},
    \end{cases}
\end{equation}
where $\text{Re}(\cdot)$ and $\text{Im}(\cdot)$ represent the real and imaginary parts of the complex input, respectively. Then, vector $\mathbf{z}^{(0)} = [{\mathbf{z}_1^{(0)}}^T,{\mathbf{z}_2^{(0)}}^T, \cdots, {\mathbf{z}_K^{(0)}}^T]^T$ is fed as an input into the subsequent CNN module. With the above preprocessing, the input vector $\mathbf{z}^{(0)}$ has a certain shift-invariance property. For example, if we shift the input vector $\mathbf z^{(0)}$ by $4B+4$ (i.e., the length of a subvector $\mathbf{z}_k^{(0)}$), we only need to shift the output vector by $1$ to obtain the same input-output relationship. With the preprocessing, a CNN can be employed to extract features of the input.

\subsection{CNN}
As shown in \figref{fig:CNN}(b), the CNN module consists of multiple convolutional hidden layers and one convolutional output layer. The input is $\mathbf{z}^{(0)}$ and the output is the symbols $\{s_k\}$, where
\begin{equation}
    s_k=\begin{cases}
    0,\text{if } x_k=-1,\\
    1, \text{if } x_k=1.
    \end{cases}
\end{equation}
We use ReLU as the activation function for the hidden layers:
\begin{equation}
    y=\text{ReLU}(x)=max(x,0),
\end{equation}
where $x \in \mathbb{R}$ is the input, and $y$ is the output of the activation function. To map the output to interval $(0,1)$, we choose the sigmoid function as the activation function for the output layer: 
\begin{equation}
    y= \text{sigmoid}(x)=\frac{1}{1+e^{-x}}.
\end{equation}
In the first convolutional layer, we use zero-padding with stride size $4B+4$, and set the filter size to $(2B+1)(4B+4)\times l_1$, where $l_1$ is the depth of the filter. That is, the $k$th output subvector $\mathbf{z}^{(1)}_k \in \mathbb{R}^{l_1}$ of the first layer is given by
\begin{equation}
    \mathbf{z}^{(1)}_k=\text{ReLU}( \mathbf{w}^{(1)}\widehat{\mathbf{z}}^{(0)}_k+\mathbf{b}^{(1)}), \label{eqn:CNN}
\end{equation}
where $\mathbf{w}^{(1)}\in \mathbb{R}^{l_1\times (2B+1)(4B+4)}$ and $\mathbf{b}^{(1)}\in \mathbb{R}^{l_1}$ are the learnable weight and bias of the first layer, and $\widehat{\mathbf{z}}^{(0)}_k = [{\mathbf{z}^{(0)}_{k-B}}^T,{\mathbf{z}^{(0)}_{k-B+1}}^T,\cdots, {\mathbf{z}^{(0)}_{k+B}}^T]^T$ with $\mathbf{z}^{(0)}_i=\mathbf{0}$ for $i<1$ or $i>K$. As such, each filter takes $2B+1$ subvectors as the input. This setting is based on the observation that each subvector is strongly correlated with $2B$ neighbouring subvectors due to the banded structure of channel $\mathbf{H}$. Hence, we propose to extract features from every $2B+1$ consecutive subvectors. Similarly, in the $i$th layer ($i>1$), the filter is performed over $2B+1$ subvectors with stride size $l_{i-1}$ and filter size $(2B+1)l_{i-1}\times l_i$. To summarize, the structure of a CNN is determined by its number of layers and the filter depth in each layer. These parameters need to be decided before training the network. As shown in simulations later, such a CNN outperforms a DNN consisting of fully-connected layers in both accuracy and complexity. 

\rmk A conventional CNN typically consists of convolutional layers as well as pooling layers and fully-connected layers. However, the fully-connected layers and pooling layers are not used in our design for the following reasons. First, a pooling layer is typically used after a convolutional layer to perform a downsampling operation along the spatial dimensions. Recall that in the proposed CNN, the filter in the convolutional layer is used to extract features for each receiving position, which means that every output of the filter is useful. Discarding features will cause performance loss. Second, the fully-connected layers involve high complexity and are also difficult to train. As shown in the simulation section, the fully-connected layers do not provide any performance gain over the convolutional layers. Hence, we have not included any pooling layers and fully-connected layers in the proposed CNN. Dropout and batch normalization are also very important components in the conventional CNN architecture. However, we have tested their performance and found that they do not provide any gain either.

\subsection{Output Postprocessing}
In the output postprocessing module, we map the output of the CNN to the estimate of the transmitted signals. Recall that we use the sigmoid function as the activation function of the output layer. As such, the output of the CNN $\hat{\mathbf{s}}$ lies in the interval $[0,1)$. Here, we use an indicator function $\mathbbm{1}(\cdot)$ to map the continuous value of output $\hat{s}_k$ to a discrete estimate of the transmitted signal $x_k$:
\begin{equation}
    \hat{x}_k = 2\mathbbm{1}(\hat{s}_k>0.5)-1.
\end{equation}
\rmk In this paper, we focus on the signal detection problem. Typically, detection and decoding are jointly considered in a communication system. The detector and the decoder iteratively exchange information (i.e., soft decisions) on the transmitted signals until convergence. The proposed CNN-based detector can be easily extend to such a iterative detecting and decoding algorithm by allowing the input preprocessing module take soft decisions of the decoder as input and allowing the output postprocessing module output soft detection decisions.
\subsection{Robustness to Different System Sizes}
DNN training typically requires a large amount of data and involves high complexity, which leads to a heavy burden on the storage and computation devices. Furthermore, the resulting DNN heavily depends on the training data, implying that a DNN has to be retrained once the system configuration changes. In this subsection, we show that our proposed CNN-based detector is robust to different system sizes in the sense that the tunable parameters do not vary with the system size $K$, and hence do not need to be retrained as long as the bandwidth $B$ does not change. Moreover, the following numerical results show that the performance of the proposed CNN-based detector is insensitive to the mismatch of the system sizes between the training set and the testing set. 

In the following experiment, we adopt three convolutional layers with depth $l_1=160, l_2=80,$ and $l_3=40$, respectively, in the proposed CNN. Each sample in the training set is independently generated with the same distribution. The cost function is the mean square error between the output $\Hat{\mathbf{s}}$ and the transmitted symbol $\mathbf{s}$. The optimization algorithm used for training is the RMSprop algorithm \cite{tieleman2012lecture} with learning rate $0.001$. We assume that the non-zero channel coefficients are independently drawn from a complex Gaussian distribution $\mathcal{CN}(0,1)$, which is widely used in communication channels with Rayleigh fading. We also assume that the transmitted signals are uniformly distributed. Moreover, we assume that the noises $\{n_k\}$ are i.i.d. drawn from $\mathcal{CN}(0,\sigma^2)$. The variance of noise is unknown and therefore it is randomly generated so that the SNR will be uniformly distributed on $[5\text{dB}, 13\text{dB}]$. This assumption allows the proposed CNN-based detector to detect over a wide range of SNR values once it is well trained.

\begin{figure}[!h]
\begin{centering}
\includegraphics[width=0.7\textwidth]{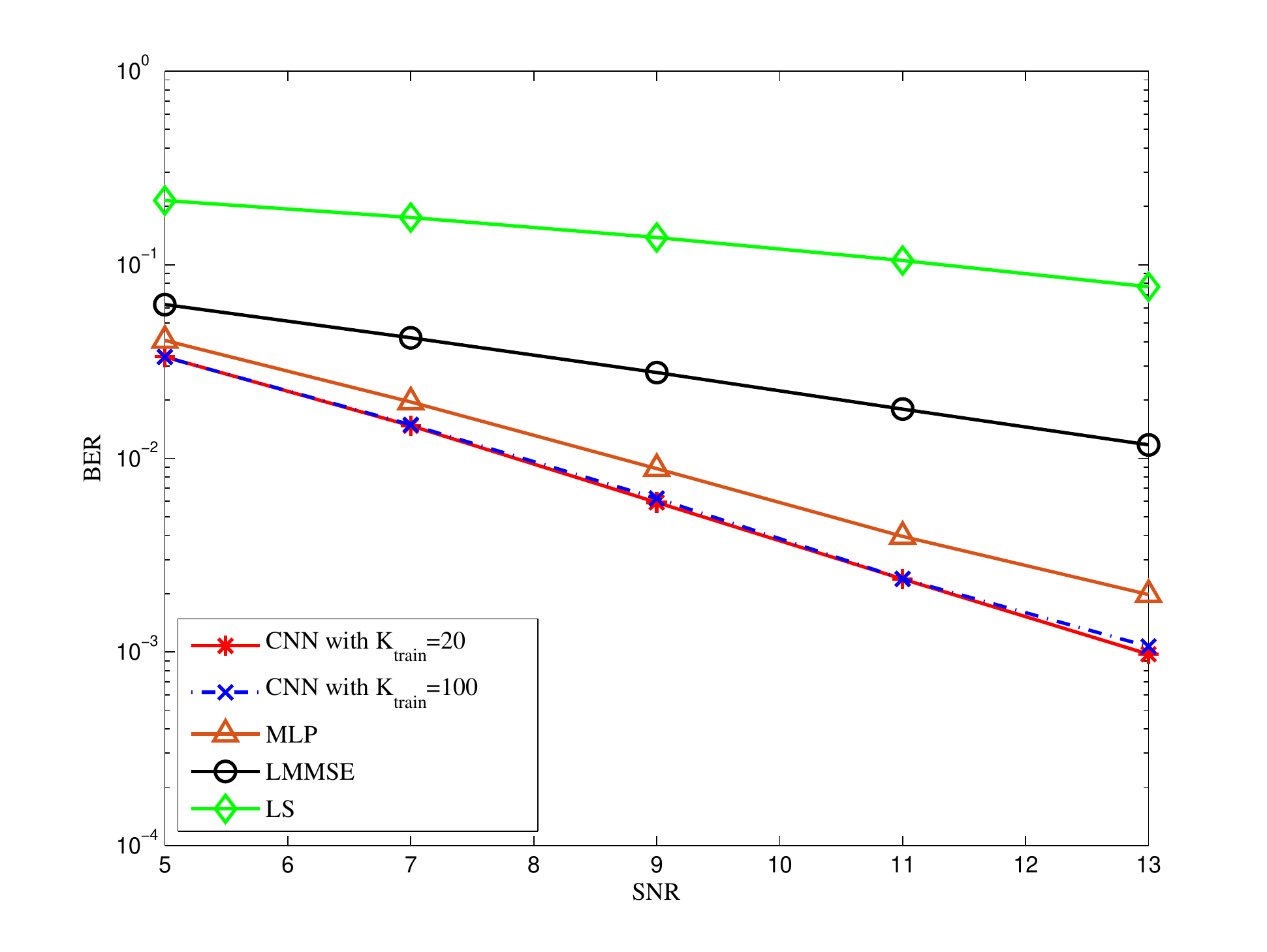}
\par\end{centering}
\centering{}\caption{Comparison of BER performance between the CNN-based detector and traditional detection algorithms with $K=20$, $B=1$, and Gaussian noise.}\label{fig:K20B1}
\end{figure}

\begin{figure}[!h]
\begin{centering}
\includegraphics[width=0.7\textwidth]{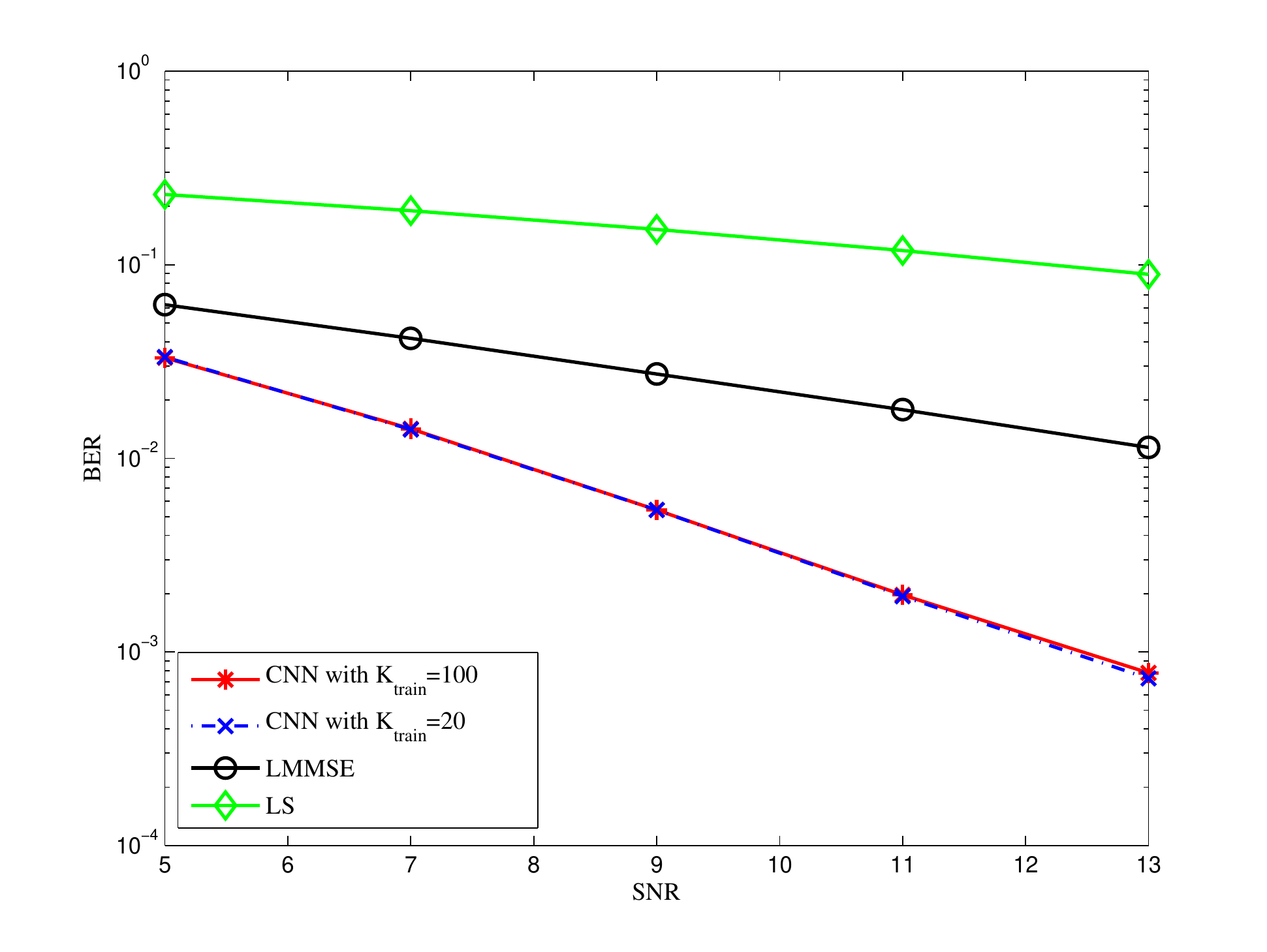}
\par\end{centering}
\centering{}\caption{Comparison of BER performance between the CNN-based detector and traditional detection algorithms with $K=100$, $B=1$, and Gaussian noise.}\label{fig:K100B1}
\end{figure}

In Fig. \ref{fig:K20B1}, we illustrate the BER curves for $K=20$ and $B=1$ with different system sizes $K_{train}$ in the training sets. The red solid curve shows the BER when the system size in the training set is the same as the size in the testing set (i.e., $K=K_{train}=20$). The blue dash-dotted curve shows the BER performance when the system sizes in the training set and the testing set are different (i.e., $K=20$ and $K_{train}=100$). As shown in Fig. \ref{fig:K20B1}, the BER gap between different training sets is negligible, implying the insensitive of the performance with respect to the training system size. Moreover, we compare the performance of the proposed CNN-based detector with three benchmark algorithms, namely, LMMSE, LS, and a DNN detector with multilayer perceptron (MLP). Compared with LMMSE and LS detectors, the proposed detector achieves more than one order of magnitude lower BER, because the CNN involves non-linearity by employing non-linear activation functions. In the MLP, we set the dimension of the output in each hidden layer to be the same that in CNN for fair comparison. That is, we use three hidden layers with $3200$, $1600$, and $800$ neurons, respectively, in the MLP. As shown in Fig. \ref{fig:K20B1}, the proposed CNN-based detector achieves a lower BER than the MLP-based detector. In addition, the MLP trained with $K_{train}=20$ cannot be applied to systems with $K\neq 20$. Meanwhile, our CNN-based detector can be applied to systems with different $K$. In Fig. \ref{fig:K100B1}, we plot the BER curves for $K=100$ and $B=1$ with training sets with $K_{train} =20$ and $K_{train}=100$. Again, we see that the BER performance of the proposed CNN-based detectors are very close to each other when the system sizes of the training sets are different. As such, we demonstrate the robustness of the proposed CNN-based detector for the mismatch in the system size. That is, once the CNN-based detector is well trained, it can be applied to systems with different sizes $K$. On the other hand, we have omitted the BER curve for a MLP-based-detector in \figref{fig:K100B1} due to the prohibitively high computational complexity and high storage requirement. For a system with $K=100$, we have to construct a MLP consisting of three hidden layers with $16000$, $8000$, and $4000$ neurons, respectively, to obtain the same output dimension with the proposed-CNN-based detector.


\section{Extension to Near-Banded Channels}
So far, we have focused on the banded linear system. However, in real applications, the channel matrix may not be a strictly banded matrix. In this section, we use two examples to show how to adjust the proposed CNN-based detector to systems with near-banded channels.
\begin{figure}[!h]
\centering
\includegraphics[scale=0.42]{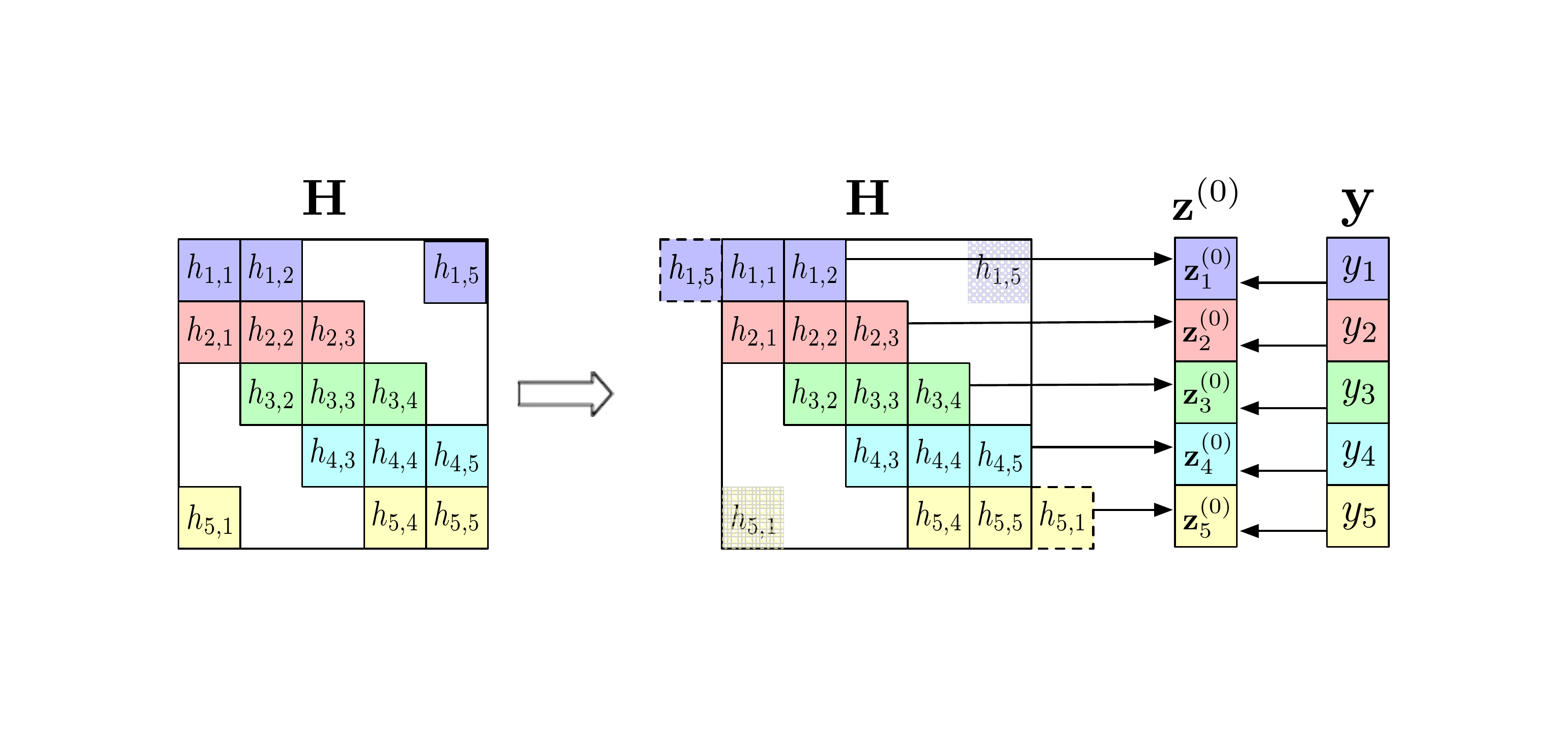}
\caption{A near-band channel matrix and its corresponding input preprocessing approach with $K=5$ and $B=1$.}\label{fig:nearband}
\end{figure}
\subsection{1-D Near-Banded System}
\begin{figure}[!h]
\centering
\includegraphics[scale=0.42]{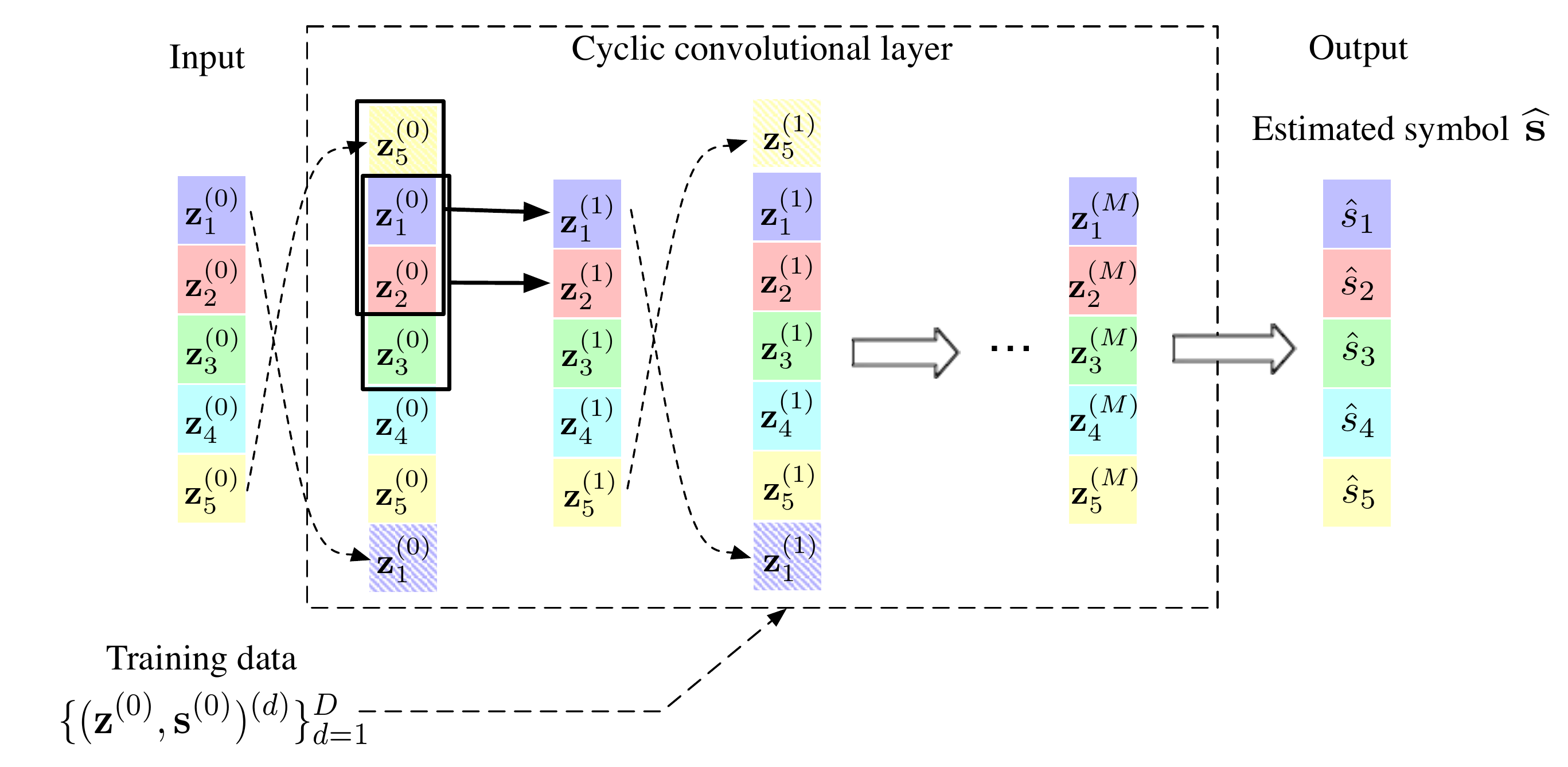}
\caption{A cyclic CNN (CCNN) architecture with $K=5$ and $B=1$.}\label{fig:CCNN}
\end{figure}
As mentioned in Section II, the channel matrix in a doubly selective OFDM system has non-zero entries in the bottom-left corner and the top-right corner of the channel matrix due to the non-negligible ICI \cite{schniter2004low}. The proposed CNN can be extended to such a near-banded system by modifying the input preprocessing module. As shown in Fig. \ref{fig:nearband}, the entries of the subvector $\mathbf{z}^{(0)}_k$ are given by 
\begin{equation}
    z_{k}^{(0)} [2i+1]\!=\! \begin{cases}
    \text{Re}(H_{k,m}), &i\leq 2B, m =1+ ((k+i-B+K-1)\mod K), \\
    \text{Re}(y_{k}), & i=2B+1,\\
    0, &\text{otherwise},
    \end{cases}
\end{equation}
and
\begin{equation}
    z_{k}^{(0)} [2i+2]\!=\!\begin{cases}
    \text{Im}(H_{k,m}), &i\leq 2B, m =1+ ((k+i-B+K-1)\mod K), \\
    \text{Im}(y_{k}), & i=2B+1,\\
    0, &\text{otherwise},
    \end{cases}
\end{equation}
where $\mod$ is the modulo operator. Then, the input vector $\mathbf{z}^{(0)}$ can be fed into the CNN proposed in Section III.C for detection. In addition, we can adjust the CNN in Section III.C to further improve the performance. Note that the subvector $z_{k}^{(0)}$ with $k\leq B$ is not only strongly correlated with $k+B-1$ neighbouring subvectors, but also strongly correlated with the subvectors $z_{m}^{(0)}$ where $m\geq K+k-B$. Here, we propose to replace the original convolutional layer with zero padding in Eqn. (\ref{eqn:CNN}) with a cyclic convolutional layer without zero padding. That is, the $k$th output subvector $\mathbf{z}^{(1)}_k \in \mathbb{R}^{l_1}$ of the first layer now becomes
\begin{equation}
    \mathbf{z}^{(1)}_k=\text{ReLU}( \mathbf{w}^{(1)}\widehat{\mathbf{z}}^{(0)}_k+\mathbf{b}^{(1)}), \label{eqn:CCNN}
\end{equation}
where $\widehat{\mathbf{z}}^{(0)}_k = [{\mathbf{z}^{(0)}_{k-B}}^T,{\mathbf{z}^{(0)}_{k-B+1}}^T,\cdots, {\mathbf{z}^{(0)}_{k+B}}^T]^T$ with $\mathbf{z}^{(0)}_i=\mathbf{z}^{(0)}_{m=1+(k+i-B+K-1)\mod K}$ for $i<1$ or $i>K$. Similarly, we can replace the $i$th layer with a cyclic convolutional layer (see \figref{fig:CCNN}). The modified detector is referred to as cyclic CNN (CCNN)-based detector. In the simulation section, we will show that both the original CNN-based detector and the CCNN-based detector perform well in near-banded systems. 
\subsection{2-D Near-Banded System}
As shown in eqn. (\ref{eqn:2D-ISI}), the received signals in a TDMR system are typically modeled by convolving the data with a 2-D spatial impulse response \cite{wu2003iterative}. The TDMR ISI system is a 2-D extension of a banded linear system studied in the previous sections. Each pair of the input element and the output element, say $(x_{j,k}, y_{j,k})$, is strongly correlated with only $2B+1$ neighbouring pairs in each dimension. Hence, we can extend the proposed CNN-based detector to a 2-D CNN-based detector for the TDMR system. We use the same activation functions for the 2-D CNN. That is, the ReLU function is used in the hidden layers, and the sigmoid function is used in the output layer. Unlike the banded linear system which has a distinct impulse response vector for each received signal, as shown in \cite{wu2003iterative}, the impulse response matrix in (\ref{eqn:2D-ISI}) is fixed for all output elements. Hence, the $(j,k)$th output subvector $\mathbf z_{j,k}^{(1)} \in \mathbb{R}^{l_1}$ of the first convolutional layer with depth $l_1$ is given by
\begin{equation}
    \mathbf{z}_{n,k}^{(1)} = \text{ReLU}\left(\sum_{m=1}^{2B+1}\sum_{l=1}^{2B+1}\mathbf{w}_{m,l}^{(1)}y_{n+m-B-1,k+l-B-1} + \sum_{m=1}^{B+1}\sum_{l=1}^{B+1}h_{m,l}\mathbf v_{m,l}^{(1)}+\mathbf{b}^{(1)}\right),
\end{equation}
where $\mathbf{w}_{m,n}^{(1)}\in \mathbb{R}^{l_1}$ and $\mathbf{v}_{m,n}^{(1)}\in \mathbf{R}^{l_1}$ are the learnable weights, and $\mathbf{b}^{(1)}\in \mathbb{R}^{l_1}$ is the bias. Then, the $(j,k)$th output subvector $\mathbf z_{j,k}^{(i)} \in \mathbb{R}^{l_k}$ of the $i$th convolutional layer is given by 
\begin{equation}
        \mathbf{z}_{n,k}^{(i)} = \text{ReLU}\left(\sum_{m=1}^{2B+1}\sum_{l=1}^{2B+1}\mathbf{w}_{m,l}^{(i)}\mathbf z_{j+m-B-1,k+l-B-1}^{(i-1)} +\mathbf{b}^{(i)}\right),
\end{equation}
with $l_k$ is the depth of the filter, $\mathbf{w}_{m,n}^{(i)}\in \mathbb{R}^{l_i\times l_{i-1}}$ and $\mathbf{b}^{(1)}\in \mathbb{R}^{l_i}$ are the learnable weight and bias. As shown in simulations, the proposed 2D-CNN-based detector can achieve a high detection accuracy with low complexity.

\section{Numerical Results}
In this section, we present simulation results to demonstrate the performance of the proposed CNN with different system settings. We will then implement CNN in practical OFDM and TDRM systems. In this section, we compare the proposed CNN-based detector with two linear benchmark algorithms: the linear minimum mean square error (LMMSE) detector and the least square (LS) detector. 
\subsection{Simulation Results}
\begin{figure}[!h]
\begin{centering}
\includegraphics[width=0.7\textwidth]{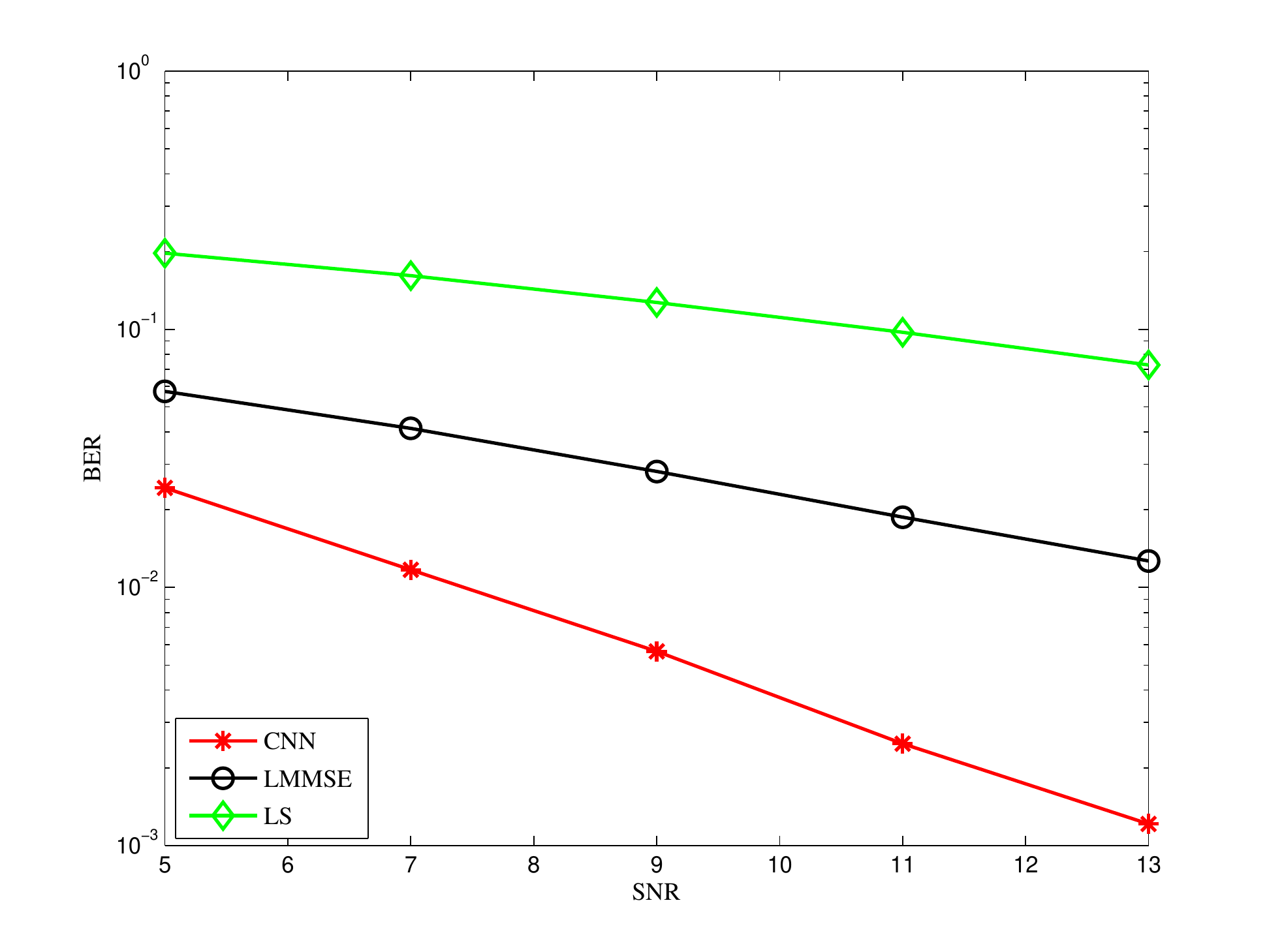}
\par\end{centering}
\centering{}\caption{Comparison of BER performance between the CNN-based detector and traditional detection algorithms in a banded system with $K=100$, $B=1$, and non-Gaussian noise.}\label{fig:BER-GM}
\end{figure}

In this subsection, we use the same CNN architecture and the same training procedure as in \figref{fig:K20B1} with the system size in the training set $K_{train}=100$. The distributions of the channel coefficients and noise are also the same as those in \figref{fig:K20B1}, except that we use a non-Gaussian noise in \figref{fig:BER-GM}.

\begin{figure}[!h]
\begin{centering}
\includegraphics[width=0.7\textwidth]{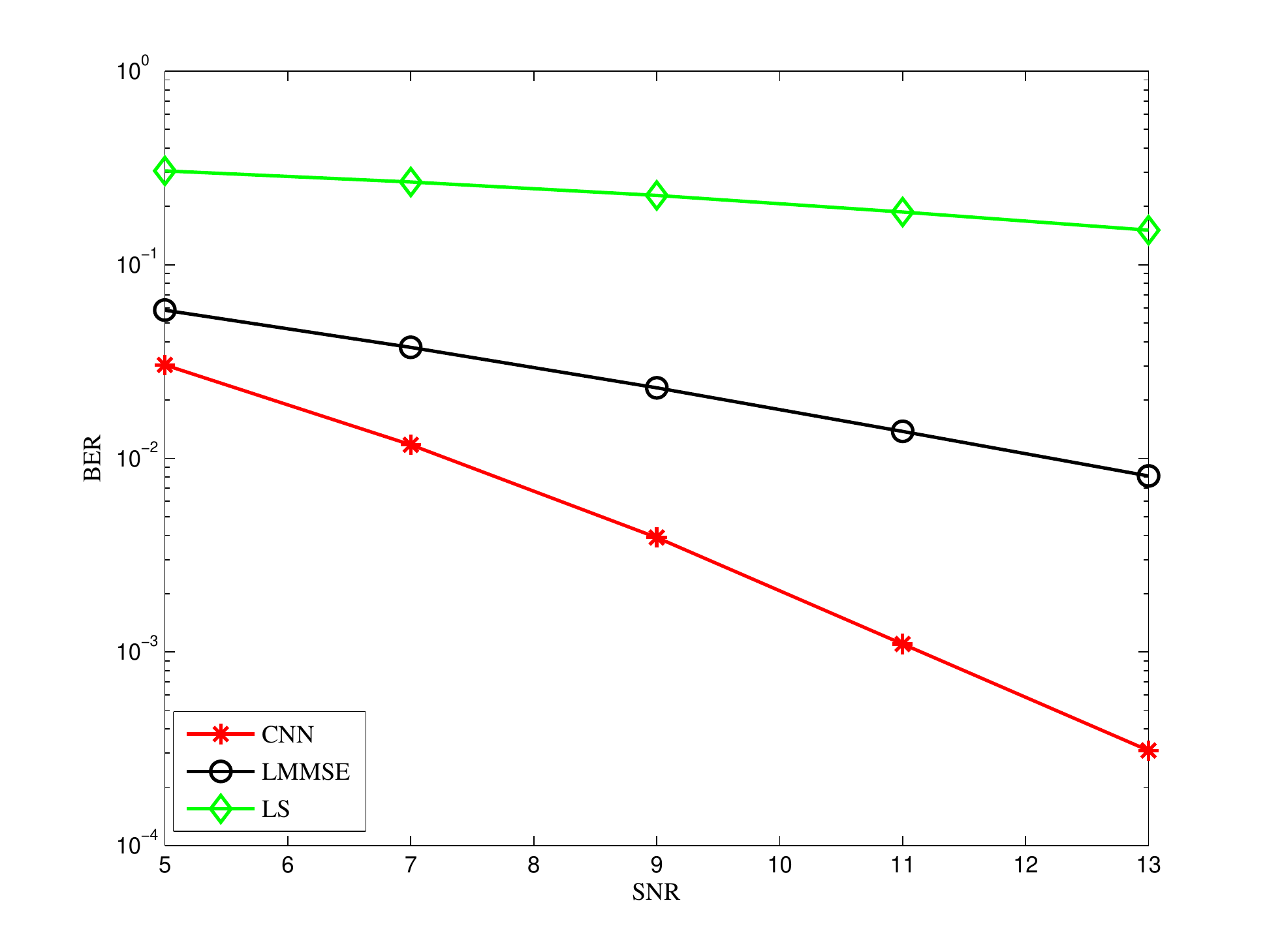}
\par\end{centering}
\centering{}\caption{Comparison of BER performance between the CNN-based detector and traditional detection algorithms in a banded system with $K=100$, $B=2$, and Gaussian noise.}\label{fig:BER}
\end{figure}

\subsubsection{Banded System}
In \figref{fig:BER}, we illustrate the BER curves for the banded linear system with $K=100$ and a wider band $B=2$. We assume that the noises $\{n_k\}$ are i.i.d. drawn from $\mathcal{CN}(0,\sigma^2)$. As shown in \figref{fig:BER}, the CNN can achieve a much lower BER than the LS and LMMSE methods. Hence, the proposed CNN-based detector can be adapted to a system with a wider band. In addition, we show that the proposed detector can be adapted to different channel and noise distributions. \figref{fig:BER-GM} plots the BER curves in a banded system with non-Gaussian noises. In particular, the noises $\{n_k\}$ are i.i.d. with each following a complex Gaussian mixture distribution:
\begin{equation}
    f(x)=0.9\mathcal{CN}(0,\sigma^2)(x)+0.1\mathcal{CN}(0,10\sigma^2)(x).\label{eqn:GM}
\end{equation}
As shown in \figref{fig:BER-GM}, the proposed detector outperforms the LS and LMMSE methods. We would like to emphasize that the Gaussian mixture noise is used only as an example of non-Gaussian noise here. The proposed CNN also work for other channel and noise distributions as long as a sufficient amount of training data with the same distributions is available.

\begin{figure}[!h]
\begin{centering}
\includegraphics[width=0.7\textwidth]{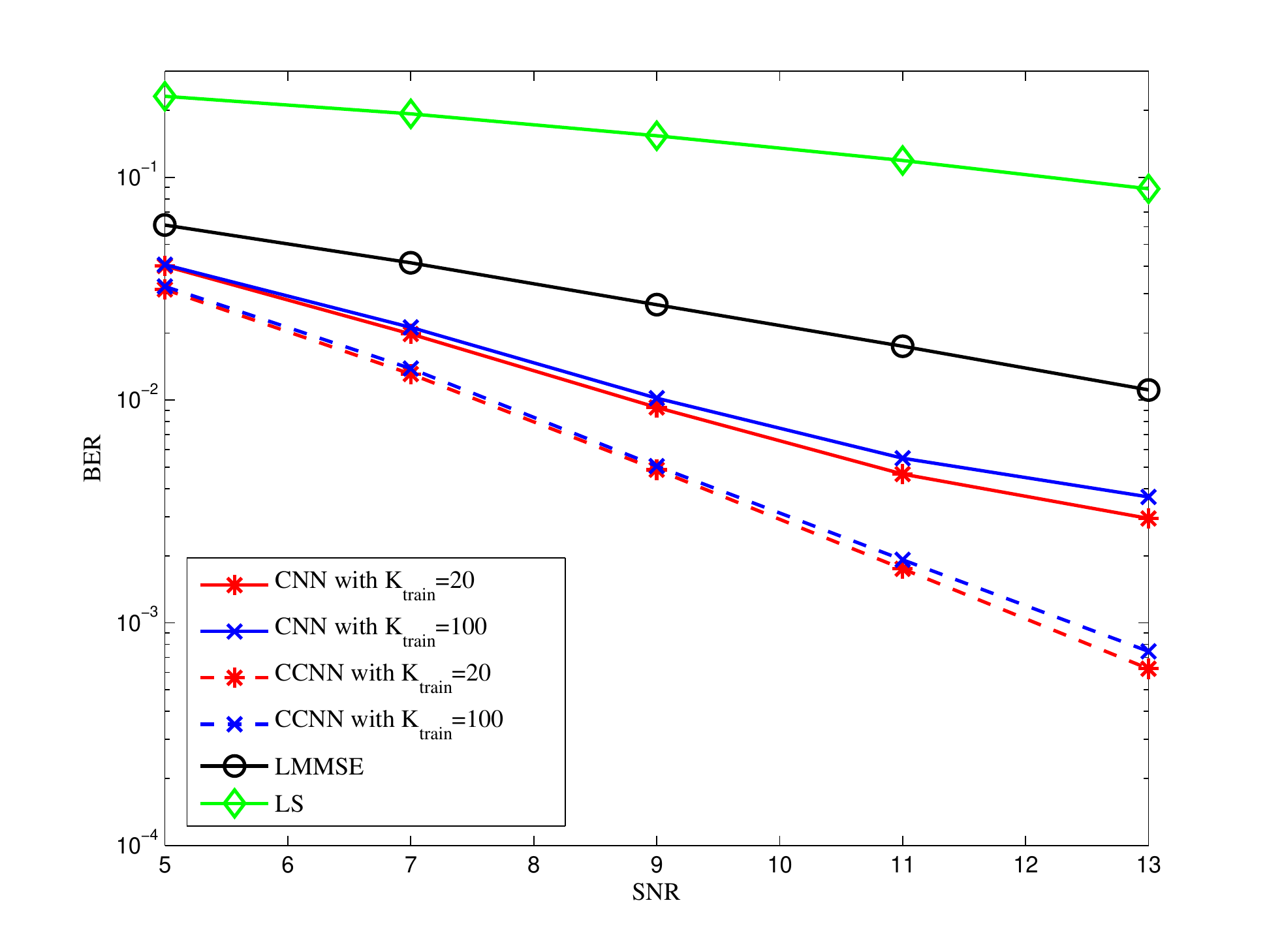}
\par\end{centering}
\centering{}\caption{Comparison of BER performance between the CNN-based detectors and traditional detection algorithms in a 1-D near-banded system with $K=20$, $B=1$, and Gaussian noise.}\label{fig:nearband1}
\end{figure}

\begin{figure}[!h]
\begin{centering}
\includegraphics[width=0.7\textwidth]{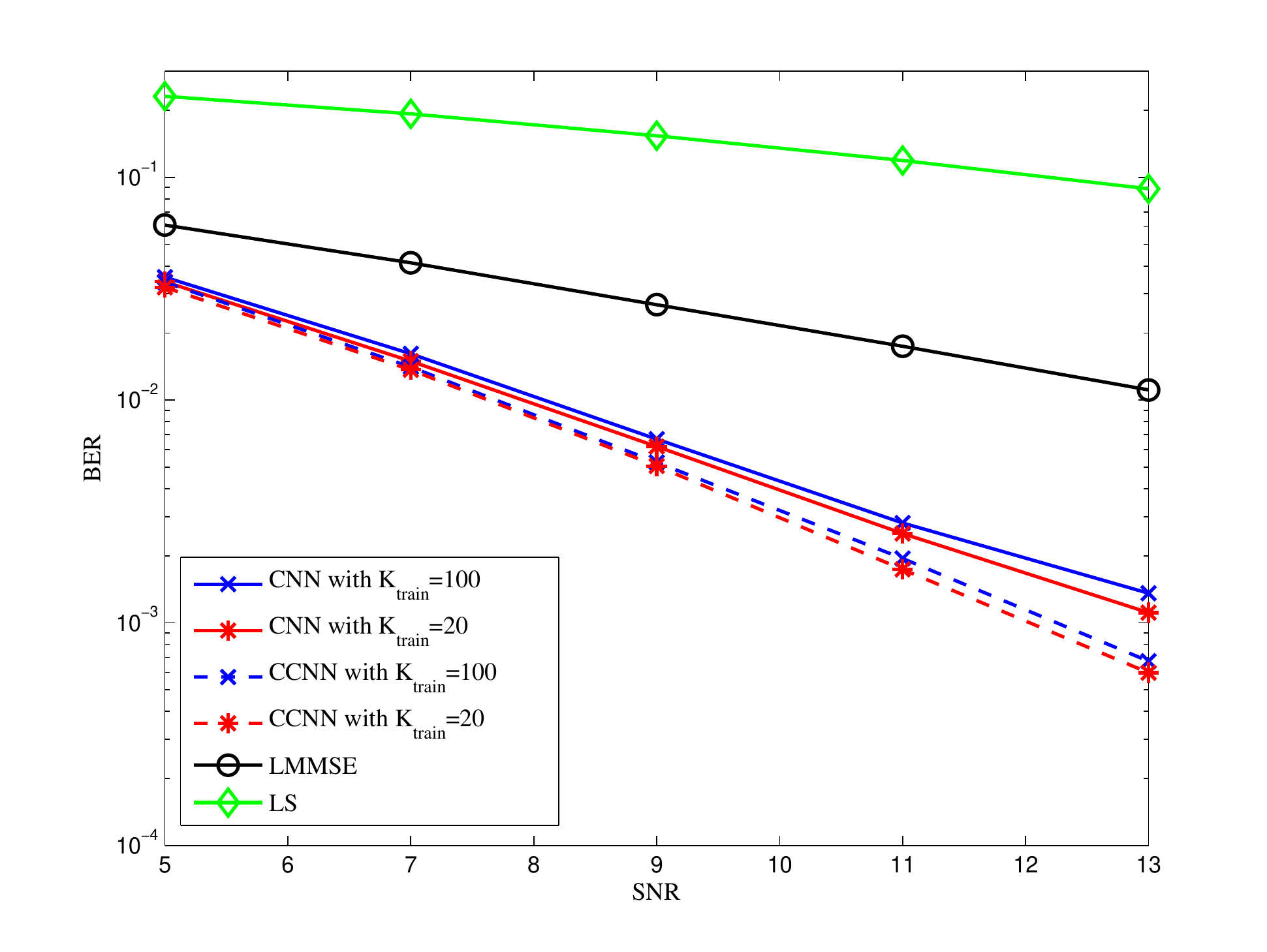}
\par\end{centering}
\centering{}\caption{Comparison of BER performance between the CNN-based detectors and traditional detection algorithms in a 1-D near-banded system with $K=100$, $B=1$, and Gaussian noise.}\label{fig:nearband2}
\end{figure}
\subsubsection{1-D Near-Banded System}
In \figref{fig:nearband1}, we illustrate the BER performance for the 1-D near-banded system with $K=20$, $B=1$, and Gaussian noise. As shown in \figref{fig:nearband1}, the proposed deep-learning-based detectors achieve a much lower BER than the LS and LMMSE methods. In addition, the CCNN-based detector performs better than the CNN-based detector. We also plot the BER curves when there is a mismatch of the system sizes in the training and testing set. The performance gap of CCNN caused by the mismatch is smaller than that of CNN. Hence, the CCNN-based detector outperforms the CNN-based detector in terms of both accuracy and robustness when the system size $K$ is small. \figref{fig:nearband2} plots the BER curves for a large system with $K=100$ and $B=1$. The performance of CCNN and CNN is very close to each other. The reason is that the effect of the non-zero entries in the bottom-left and top-right corners is negligible when $K\gg B$. 

\subsection{OFDM Systems}
In this subsection, we implement the proposed CNN in two OFDM systems: the underwater acoustic system in \cite{berger2010sparse} with a strictly banded channel matrix, and the time- and frequency-selective (or doubly selective) OFDM system in \cite{schniter2004low} with a near-banded channel matrix. 
\subsubsection{Underwater Acoustic Systems}
\begin{figure}[!h]
\begin{centering}
\includegraphics[width=0.7\textwidth]{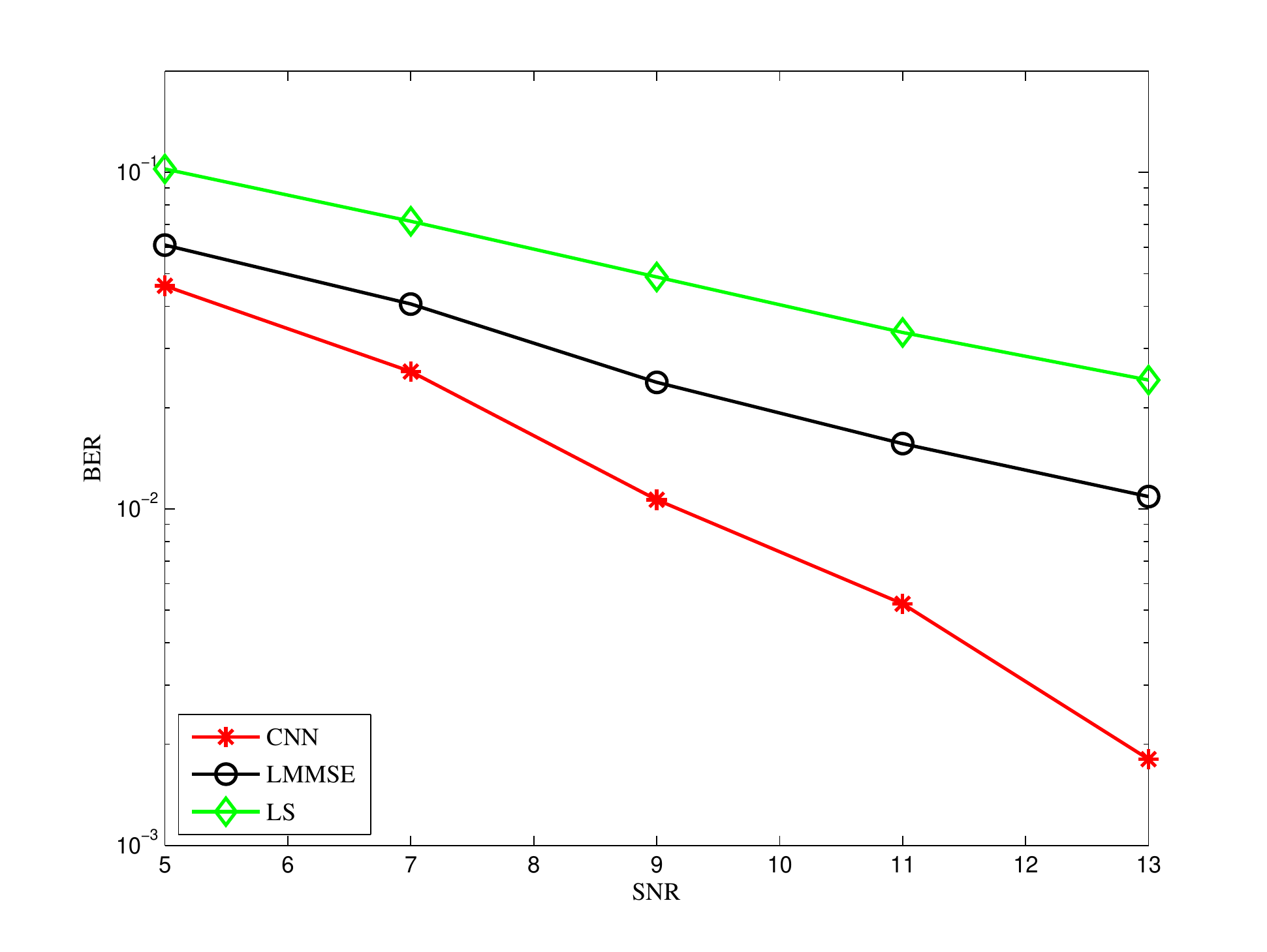}
\par\end{centering}
\centering{}\caption{Comparison of BER performance between the CNN-based detector and traditional detection algorithms in an underwater acoustic system with $K=1024$ and $B=1$.}\label{fig:OFDM_1}
\end{figure}
In \cite{berger2010sparse}, an underwater acoustic time-varying multipath channel consists of $N_p$ discrete paths is considered. The channel impulse response is defined as 
\begin{equation}
    h(\tau, t)=\sum_{p=1}^{N_p}A_p(t)\delta(\tau-\tau_p(t)),
\end{equation}
with $A_p(t)$ and $\tau_p(t)$ being the amplitude and delay of the $p$th path. During an OFDM symbol, the time variation of the path delays is approximated by
\begin{equation}
    \tau_p(t)\approx \tau_p-a_pt,
\end{equation}
and the path amplitudes are assumed constant $A_p(t)=A_p$. Moreover, zero padding is employed in the system, where $T$ denotes the OFDM symbol duration, and the subcarrier spacing is $1/T$. The $k$th subcarrier is at frequency 
\begin{equation}
    f_k=f_c+k/T, k=-K/2,\cdots,K/2-1,
\end{equation}
where $f_c$ is the carrier frequency. Then, by using the banded assumption in \cite{berger2010sparse}, the fast Fourier transform (FFT) output on the $k$th subcarrier is
\begin{equation}
    y_k=\sum_{b=-B}^{B}H_{k,k+b}x_{k+b}+n_k,
\end{equation}
where 
\begin{equation}
    H_{k,m}=\sum_{p=1}^{N_p}\frac{A_p}{1+a_p}e^{-j2\pi f_k\tau_p'}\varrho_{k,m}^{(p)},
\end{equation}
with $\tau_p'=\frac{\tau_p}{1+a_p}$ and
\begin{equation}
    \varrho_{k,m}^{(p)}=\frac{\sin(\pi \beta_{k,m}^{(p)}T)}{\pi \beta_{k,m}^{(p)}T}e^{j\pi \beta_{k,m}^{(p)}T}
\end{equation}
\begin{equation}
    \beta_{k,m}^{(p)} = (m-k)\frac{1}{T} + \frac{a_pf_k}{1+a_p}.
\end{equation}
\figref{fig:OFDM_1} illustrates the BER performance of the proposed CNN-based detector. The simulation settings are the same as those in \cite{berger2010sparse}, with symbol duration $T=104.86$ms, carrier frequency $f_c=13$kHz, $K=1024$ subcarriers, and $N_p=15$ discrete paths. Each path has a separate zero-mean Doppler rate, which is drawn from a uniform distribution with standard deviation of $\sigma_v=0.3$m$/$s. We also assume that the transmitted symbols are i.i.d. BPSK symbols and the noise follows a i.i.d. Gaussian distribution. As shown in \figref{fig:OFDM_1}, the CNN-based detector significantly outperforms both the LMMSE and LS detectors. 

\subsubsection{Doubly Selective OFDM Systems}

\begin{figure}[!h]
\begin{centering}
\includegraphics[width=0.7\textwidth]{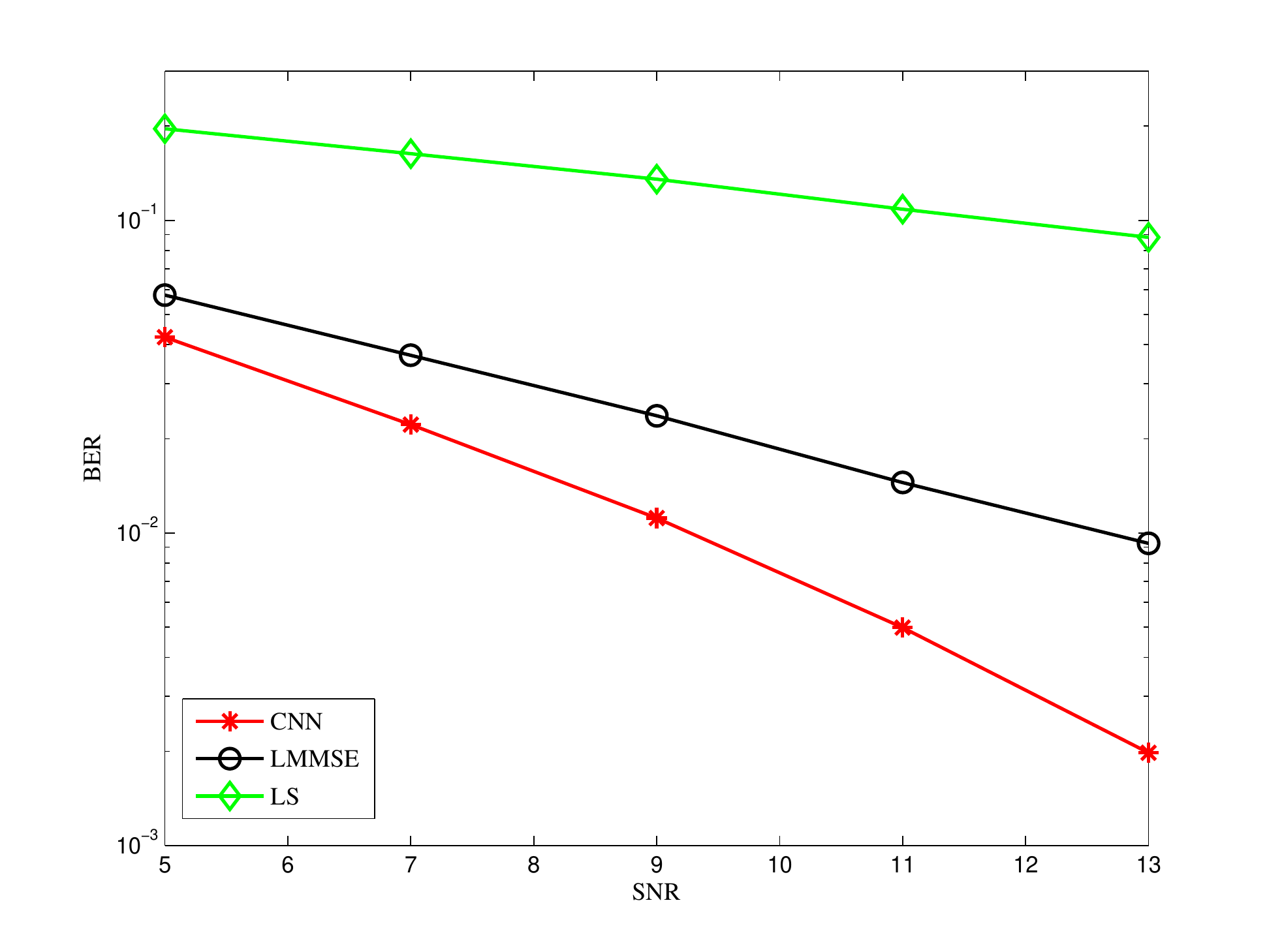}
\par\end{centering}
\centering{}\caption{Comparison of BER performance between the CNN-based detector and traditional detection algorithms in a doubly selective OFDM system with $K=128$ and $B=1$.}\label{fig:OFDM_2}
\end{figure}
In \cite{schniter2004low}, an OFDM system over a noisy multipath channel is considered. The multipath channel is modeled by the time-variant discrete impulse response $h_{n,l}^{(tl)}, \forall{l\leq N_h}$, defined as the time-n response to an impulse applied at time $n-l$. $N_h$ is the maximum delay spread. Incorporating the transmitted symbols $\mathbf{x}\in \mathbb{C}^{K}$ with a cyclic prefix of length $N_p\geq N_h$, the FFT output on the $k$th subcarrier is
\begin{equation}
    y_k = \sum_{m=1}^K H_{k,m}x_m+n_k,
\end{equation}
where 
\begin{equation}
    H_{k,m}=\frac{1}{K} \sum_{n=0}^{K-1}\sum_{l=0}^{N_h}h_{n,l}^{(tl)}e^{-j\frac{2\pi}{K}(lm+(k-m)n)}.
\end{equation}
In this paper, we use BPSK as the modulation method, i.e., $\mathbf{x}\in \{\pm 1\}^K$. As shown in \cite{schniter2004low}, in a typical wide-sense stationary uncorrelated scattering (WSSUS) model \cite{jakes1994microwave}, $H_{k,m}\approx 0$, if $B<|k-m|<K-B$, where $B\geq \left \lceil f_dK \right \rceil$, and $f_d$ is the maximum Doppler frequency normalized to the signaling rate. As such, the doubly selective OFDM system can be modeled as the 1-D near-banded system in eqn (\ref{eqn:nearband}). In \figref{fig:OFDM_2}, we plot the BER curve for the CNN-based detector proposed in Section IV.A with $k=128$, $f_d=0.005$, and $B=1$. Realizations of $h(n,l)$ are generated following the parameter settings in \cite{schniter2004low}. That is, we assume that the transmitted symbols are i.i.d. BPSK symbols, the noise is an AWGN noise, and the channel is an energy-preserving WSSUS Rayleigh-fading channel with variance $\sigma_l^2=N_h^{-1}$, where $N_h=K/4$. Through \figref{fig:OFDM_2}, we observe that the proposed CNN-based detector achieves a much lower BER than the linear LMMSE and LS detectors. That is, the CNN-based detector can be easily extended to a system with a near-banded channel. As shown in \figref{fig:nearband2}, when $K\gg B$, the BER performances of CNN and CCNN are very close to each other. Hence, we omit the BER curve of the CCNN-based detector in \figref{fig:OFDM_2}.
\subsection{TDMR Systems}
\begin{figure}[!h]
\begin{centering}
\includegraphics[width=0.7\textwidth]{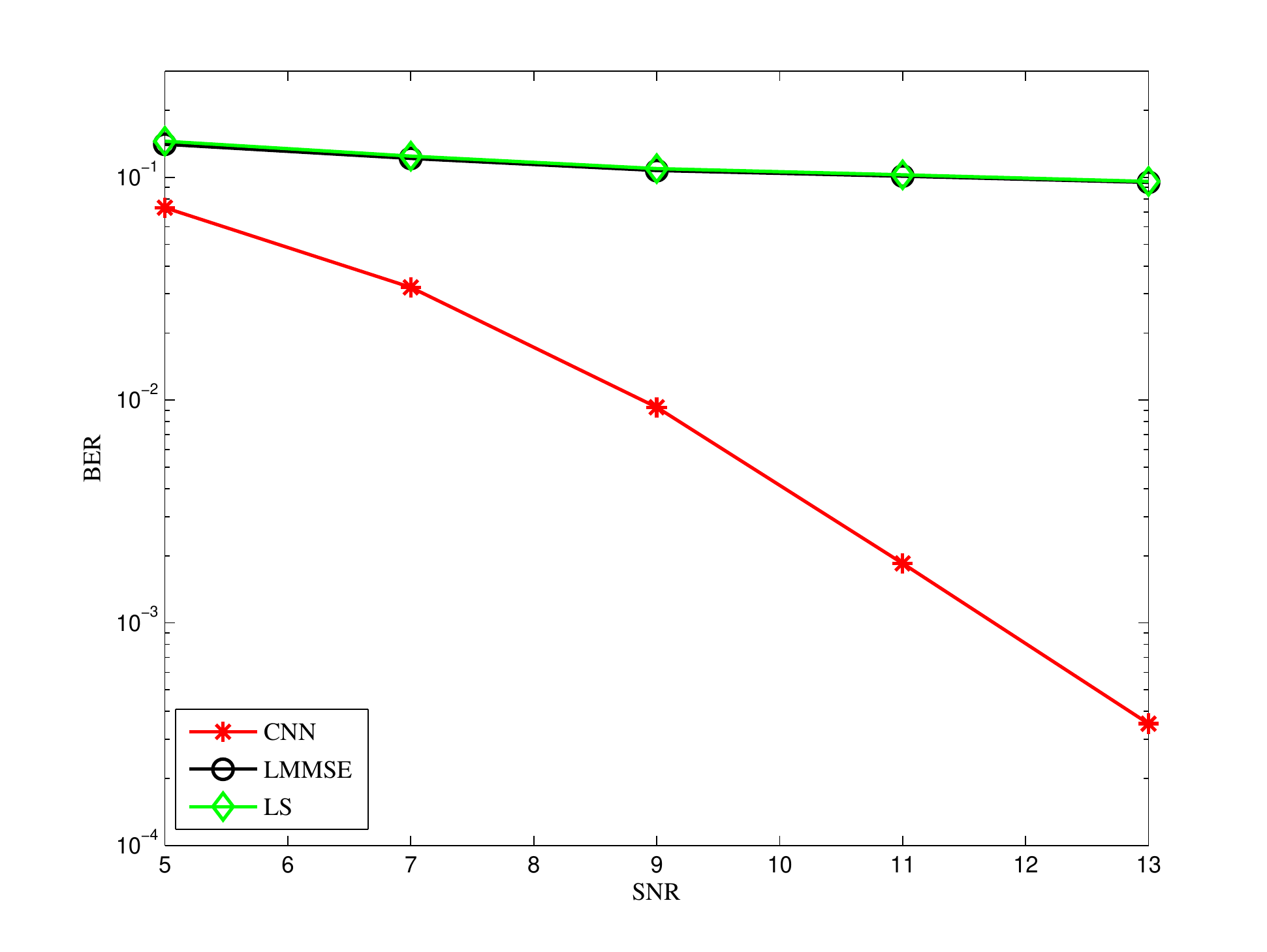}
\par\end{centering}
\centering{}\caption{BER performance of the CNN-based detector with $N=200, K=100$ and Gaussian noise.}\label{fig:TDMR}
\end{figure}

In this subsection, we show the performance of the proposed CNN-based detector in TDMR systems. The system model is given in Eqn. (\ref{eqn:2D-ISI}). We assume that $B=1$ and the channel coefficients are independently drawn from a Gaussian distribution. Fig. \ref{fig:TDMR} shows the BER performance of the proposed CNN-based detector and the 2D-LMMSE \cite{chugg1997two}, 2D-LS methods with $K=100$ and $N=200$. The BER of CNN is much lower than those of 2D-LMMSE and 2D-LS. Note that 2D-LMMSE and 2D-LS perform very badly since they detect signals only based on a very limited neighbourhood of the corresponding received signals. Extending the 2D-LMMSE and 2D-LS to traditional LMMSE and LS that take all the received signals as input will lead to prohibitively high computational complexity due to the large dimension of the TDMR system. Once again, we want to emphasize here that similar to the 1-D systems, the proposed CNN-based detector can be applied to 2-D systems with different sizes as long as the underlying distributions of the channel are the same.

\section{Conclusions and Future Work}
In this paper, we studied the use of deep learning in a signal detection problem. We focused on a specific communication system where the channel matrix is a banded matrix. We proposed a novel CNN architecture, which can achieve both high accuracy and robustness. Through simulations, we observed that the proposed CNN significantly outperform the LMMSE and LS detector in terms of BER and computational time. In addition, we showed that CNN performs better than conventional DNNs in terms of both BER and computational time, and is more suitable to large-system applications. Furthermore, we show that the proposed CNN is robust to different system sizes. The CNN-based detector also demonstrate good performance in practical systems, including doubly selective OFDM systems and TDMR systems. This work shows a great potential of deep learning, particularly CNN, for signal detection in complicated communication environments. 

The design of the CNN-based detector is inspired by existing iterative detection algorithms especially belief propagation. In a belief propagation algorithm, the probabilistic messages of a transmitted signal are updated based on the local messages of neighbouring transmitted signals, where two transmitted signals are neighbouring to each other if the corresponding transmitters share the same receiving position. In a banded linear system, two transmitted signals are neighbouring to each other if their indices are consecutive. To detect a signal, we only need to share messages between consecutive signals. In this paper, instead of calculating the messages based on the probabilistic model, we train a CNN to extract and share messages between consecutive signals. In other words, the proposed CNN-based detector unfolds the procedure of iterative algorithms, and imitates the updating rule based on real-world data. Hence, the proposed CNN-based detector can be adapted to an arbitrary communication system in which each transmitted signal only has a small number of neighbouring signals. Moreover, notice that in communication systems, a lot of problems, such as resource management, user association, and power control, can be well solved by iterative algorithms. Our proposed CNN-based architecture can be generalized to solve these problems as long as the iterative algorithms update the messages based on local information in the neighbourhood. We will leave such extension of the CNN-based architecture for future work.

\bibliographystyle{ieeetran}
\bibliography{database}

\end{document}